\documentclass[a4paper, 11pt]{article}
\usepackage[utf8]{inputenc}
\usepackage[ngermanb, english]{babel}
\usepackage{amsmath}
\usepackage{amssymb}
\usepackage{amsthm}
\usepackage{euscript}
\usepackage{tikz-cd}
\usepackage{bm}
\usepackage{commath}
\usepackage{mathrsfs}
\usepackage{mathtools}
\usepackage{slashed}
\usepackage{tensor}
\usepackage{parskip}
\usepackage{epstopdf}
\usepackage{graphicx}
\usepackage{hhline}
\usepackage{accsupp}
\usepackage[nottoc]{tocbibind}
\usepackage{babelbib}
\usepackage[top=2.5cm, left=2.5cm, right=2.5cm, bottom=2.5cm]{geometry}
\usepackage{hyperref}

\theoremstyle{plain}
\newtheorem{thm}{Theorem}[section]
\newtheorem{lem}[thm]{Lemma}
\newtheorem{prop}[thm]{Proposition}

\newtheorem{col}[thm]{Corollary}

\theoremstyle{definition}
\newtheorem{defn}[thm]{Definition}
\newtheorem{con}[thm]{Convention}

\theoremstyle{remark}
\newtheorem{rem}[thm]{Remark}

\newcommand{\Q}{{\boldsymbol{\mathrm{Q}}}}
\newcommand{\LQ}{\mathcal{L}_{\Q}}
\newcommand{\TQ}{\mathcal{T}_{\Q}}

\newcommand{\imaginary}{\mathrm{i}}
\newcommand{\gcoupling}{\varkappa}
\newcommand{\lindeDonder}{\mathscr{D}^{(1)}}
\newcommand{\Lorenz}{\mathscr{L}}
\newcommand{\textfrac}[2]{#1 / #2}
\newcommand{\dt}[1]{\operatorname{Det} \left ( #1 \right )}

\newcommand{\setbig}[1]{\bigl \{ #1 \bigr \}}
\newcommand{\cgreen}[1]{\vcenter{\hbox{\includegraphics[width=\cgreenlength]{#1}}}}
\newcommand{\tcgreen}[1]{\vcenter{\hbox{\includegraphics[width=0.75\cgreenlength]{#1}}}}
\newcommand{\enter}{\vspace{\baselineskip}}
\newcommand{\OS}{\mathcal{O}}
\newcommand{\id}{\operatorname{Id}}
\newcommand{\mettransstr}{E}
\newcommand{\gtym}{\mathbbit{\mathfrak{g}}}
\newcommand{\gtymred}{\mathbbit{\mathfrak{g}}^\prime}
\newcommand{\gtymredos}{\mathbbit{\mathfrak{g}}^\prime_\OS}
\newcommand{\gfym}{\mathbbit{\mathfrak{f}}}
\newcommand{\gtgr}{\mathbbit{\mathfrak{G}}}
\newcommand{\gtgrred}{\mathbbit{\mathfrak{G}}^\prime}
\newcommand{\gfgr}{\mathbbit{\mathfrak{F}}}

\newcommand*{\cevss}[1]{\, \reflectbox{\ensuremath{\scriptstyle \vec{\reflectbox{\ensuremath{\! \scriptstyle #1}}}}}}
\newcommand{\usc}{%
	\BeginAccSupp{method=hex,unicode,ActualText=005F}%
	\raisebox{-0.2\baselineskip}[0pt][0pt]{-}%
	\EndAccSupp{}%
}

\newsavebox{\foobox}
\newcommand{\italicbox}[2][.25]
{%
	\mbox
	{%
		\sbox{\foobox}{#2}%
		\hskip\wd\foobox
		\pdfsave
		\pdfsetmatrix{1 0 #1 1}%
		\llap{\usebox{\foobox}}%
		\pdfrestore
	}%
}

\makeatletter
\newcommand{\mathbbit}[1]{\mathpalette\mathbbit@{#1}}
\newcommand{\mathbbit@}[2]{\mathord{\mkern-0.5mu\italicbox{$\m@th#1\mathbb{#2}$}\mkern2.5mu}}
\makeatother

\makeatletter
\newcommand{\subalign}[1]{%
	\vcenter{%
		\Let@ \restore@math@cr \default@tag
		\baselineskip\fontdimen10 \scriptfont\tw@
		\advance\baselineskip\fontdimen12 \scriptfont\tw@
		\lineskip\thr@@\fontdimen8 \scriptfont\thr@@
		\lineskiplimit\lineskip
		\ialign{\hfil$\m@th\scriptstyle##$&$\m@th{\scriptstyle }##$\crcr
			#1\crcr
		}%
	}
}
\makeatother

\providecommand{\sectionref}[1]{Section~\ref{#1}}

\providecommand{\ssecref}[1]{Subsection~\ref{#1}}

\providecommand{\appendixref}[1]{Appendix~\ref{#1}}
\providecommand{\eqnref}[1]{Equation~\eqref{#1}}

\providecommand{\eqnsref}[1]{Equations~\eqref{#1}}
\providecommand{\eqnsaref}[2]{Equations~\eqref{#1} and \eqref{#2}}
\providecommand{\eqnssaref}[3]{Equations~\eqref{#1}, \eqref{#2} and \eqref{#3}}

\providecommand{\propsaref}[2]{Propositions~\ref{#1} and \ref{#2}}

\providecommand{\conref}[1]{Convention~\ref{#1}}

\providecommand{\defnsaref}[2]{Definitions~\ref{#1} and \ref{#2}}
\providecommand{\colsaref}[2]{Corollaries~\ref{#1} and \ref{#2}}

\providecommand{\lemsaref}[2]{Lemmata~\ref{#1} and \ref{#2}}
\providecommand{\thmsaref}[2]{Theoremata~\ref{#1} and \ref{#2}}

\providecommand{\remsaref}[2]{Remarks~\ref{#1} and \ref{#2}}

\newlength{\graphlength}
\newlength{\cgreenlength}
\title{\textsc{Transversality in the Coupling of Gravity\\to Gauge Theories}}
\author{David Prinz\footnote{Department of Mathematics and Department of Physics at Humboldt University of Berlin and Department of Mathematics at University of Potsdam; prinz@\{math.hu-berlin.de, physik.hu-berlin.de, math.uni-potsdam.de\}}}
\date{August 30, 2026}

\begin{document}

\maketitle

\begin{abstract}
	We consider (effective) Quantum General Relativity coupled to the Standard Model (QGR-SM) and study its transversality. To this end, we provide all propagator and three-valent vertex Feynman rules. Then, we recall the longitudinal, identical and transversal projection operators for Quantum Yang--Mills theory with a Lorenz gauge fixing. Next, we introduce the corresponding projection operators for (effective) Quantum General Relativity with a de Donder gauge fixing. With this setup, we recall several immediate identities for Quantum Yang--Mills theories and introduce their involved counterparts in (effective) Quantum General Relativity: This includes decompositions of the longitudinal projection tensors into products of linearized gauge transformation contractions and linearized gauge fixing projections. Additionally, we decompose the corresponding gauge boson and graviton propagators in terms of their transversal structure and relate them to their corresponding ghost propagators. Furthermore, we study all three-valent gravity-matter vertex contraction identities and show that they are satisfied for QGR-SM. Finally, we introduce the notion of an \emph{optimal gauge fixing} as the natural choice for any given gauge theory: In particular, we find that this is the de Donder gauge fixing in General Relativity and the Lorenz gauge fixing in Yang--Mills theory. In addition, we provide the derived Feynman rules in FORM code in the appendix and in a supplementary file.
\end{abstract}

\section{Introduction} \label{sec:introduction}

The quantization of gauge theories introduces several new challenges: Most notably the necessity to choose a gauge fixing in order to calculate the propagator. This is due to the fact that the equations of motion of the gauge field determine only the evolution of its horizontal (i.e.\ physical) degrees of freedom. The vertical (i.e.\ gauge) degrees of freedom are unconstrained due to the gauge invariance of the theory. This is an obstruction to calculating the propagator of the gauge field, as it is given as the inverse of the differential operator of the corresponding quadratic monomial. Thus, to be able to invert this differential operator, a gauge fixing term needs to be added, such that said differential operator obtains full rank and becomes invertible. This choice has important consequences for the corresponding Quantum Field Theory, as it introduces the notion of longitudinal and transversal polarizations for the external states \cite{Cvitanovic}. These degrees of freedom, which characterize propagating gauge fields, become especially important when Feynman graphs are considered: This is due to the fact that only the transversal polarizations are physical. Thus, physically consistent theories should be such that unphysical (longitudinal) degrees of freedom decouple from physical (transversal) ones. It turns out that this can be achieved by introducing ghost fields and the addition of the corresponding additional Feynman integrals. Ghost fields, at least in the original Faddeev--Popov construction \cite{Faddeev_Popov}, are constructed to satisfy residual gauge transformations as equations of motion. With this setup, it has been shown that Quantum Yang--Mills theory is indeed transversal \cite{Ward,Takahashi,tHooft,Taylor,Slavnov}. Crucially, this question is also directly linked to renormalization: Specifically, the transversal and longitudinal parts of the propagator can be renormalized individually by introducing a \(Z\)-factor for the gauge fixing parameter \cite{Prinz_3}. This then relates the counterterms of longitudinal gauge bosons with their corresponding ghosts via the so-called \emph{Slavnov--Taylor identities}.

The situation of (effective) Quantum General Relativity is much more subtle, due to the non-renormalizable nature of the theory: Specifically, it requires a \emph{UV completion} such that the longitudinal and transversal contributions of the counterterms match the corresponding integrands. Perturbative Quantum Gravity is already a well-studied field: Notably, it turns out that one-loop divergences can be absorbed via a local field redefinition in four dimensions of spacetime using the background field method \cite{tHooft_Veltman_OLD,Deser_Tsao_Nieuwenhuizen,Abbott_BFM,Donoghue}. For two-loop divergences, this turns out to be only true for non-local field redefinitions \cite{Goroff_Sagnotti_TLD,vdVen_TLD,Krasnov,Bern_Chi_Dixon_Edison_TLD}. Nevertheless, we believe that further progress can be made using modern methods, such as the Connes--Kreimer renormalization framework \cite{Kreimer_Hopf_Algebra,Connes_Kreimer_0} and the Batalin--Vilkovisky quantization formalism \cite{Batalin_Vilkovisky_1,Batalin_Vilkovisky_2}: This includes the \emph{Corolla polynomial}, which is a third graph polynomial in addition to the two Symanzik polynomials that relates the parametric representation of \(\phi^3_4\) scalar field theory amplitudes to Quantum Yang--Mills theory amplitudes \cite{Kreimer_Sars_vSuijlekom,Kreimer_Yeats,Prinz_1,Kreimer_Corolla}. In particular, the application of Connes--Kreimer theory to perturbative Quantum Gravity allows a precise specification of the obstructions for renormalization \cite{Kreimer_QG1,Kreimer_QG2}, which turn out to be gravitational Slavnov--Taylor identities for the UV completed theory \cite{Prinz_3,Prinz_9}. In this article, we study the transversality of (effective) Quantum General Relativity coupled to the Standard Model: Specifically, we aim to complement the classical literature in the following ways: First, we employ the de Donder gauge fixing with a general gauge fixing parameter \(\zeta\), while most of the literature specializes directly to the simpler Feynman gauge \(\zeta = 1\). This is relevant for renormalization, since we generally need to renormalize the gauge fixing parameter by introducing a corresponding \(Z\)-factor, i.e.\ \(\zeta \rightsquigarrow Z_\zeta \zeta\), cf.\ \cite{Prinz_3,Gracey_Kissler_Kreimer}. In addition, we obtain valuable insight into the involved tensor structure of (effective) Quantum General Relativity and are able to express the lengthy tensor expressions using a set of tensors with specific geometric interpretations, cf.\ \cite{Prinz_2,Prinz_4}. This is relevant for the ghost construction, as the ghost propagator should also be rescaled via the gauge fixing parameter \(\zeta\), such that the ghost propagator vanishes in the transversal Landau gauge \(\zeta = 0\), cf.\ \cite{Prinz_5,Prinz_6}. Furthermore, we also include the corresponding graviton-ghost and graviton-matter identities, which --- to the best of the author's knowledge --- have not been published elsewhere. Moreover, we relate the metric decomposition \(g_{\mu \nu} \equiv \eta_{\mu \nu} + \gcoupling h_{\mu \nu}\) to the metric density decomposition \(\sqrt{- \dt{g}} g^{\mu \nu} \equiv \eta^{\mu \nu} + \gcoupling \boldsymbol{\phi}^{\mu \nu}\) of Goldberg \cite{Goldberg} and Capper et al.\ \cite{Capper_Leibbrandt_Ramon-Medrano,Capper_Medrano,Capper_Namazie}: In fact, we find that the two respective propagators are dual to each other with respect to a specific metric on two-tensors. Finally, we emphasize the corresponding discussion for a non-vanishing cosmological constant \(\Lambda\) in \cite{Prinz_8} and highlight the well-written overview on different constructions and achievements within the field \cite{Burgess}.

More precisely, we consider (effective) Quantum General Relativity coupled to the Standard Model with the sign choices listed in \conref{con:sign_choices}, given via the following Lagrange density:
\begin{equation}
	\mathcal{L}_\text{QGR-SM} \coloneq \mathcal{L}_\text{QGR} + \mathcal{L}_\text{QYM} + \mathcal{L}_\text{Matter}
\end{equation}
Here, \(\mathcal{L}_\text{QGR}\) is the Lagrange density for (effective) Quantum General Relativity, \(\mathcal{L}_\text{QYM}\) is the Lagrange density for Quantum Yang--Mills theory and finally \(\mathcal{L}_\text{Matter}\) is the Lagrange density for the matter fields in the Standard Model.

Specifically, the setup for (effective) Quantum General Relativity is given as follows:
\begin{subequations}
\begin{align}
	\mathcal{L}_\text{QGR} & \coloneq \mathcal{L}_\text{GR} + \mathcal{L}_\text{GR-GF-Ghost}
	\intertext{with the Einstein--Hilbert Lagrange density}
	\mathcal{L}_\text{GR} & \coloneq - \frac{1}{2 \gcoupling^2} R \dif V_g
	\intertext{and the symmetric gauge fixing and ghost Lagrange density, as derived in \cite{Prinz_6},}
	\begin{split}
		\mathcal{L}_\text{GR-GF-Ghost} & \coloneq \frac{1}{2 \zeta} \left ( \frac{1}{2 \gcoupling^2} \eta^{\mu \nu} \lindeDonder_\mu \lindeDonder_\nu + \eta^{\mu \nu} \bigl ( \partial_\mu \overline{C}^\rho \bigr ) \bigl ( \partial_\nu C_\rho \bigr ) \right ) \dif V_\eta \\
		& \phantom{\coloneq} + \frac{1}{4} \eta^{\mu \nu} \left ( \bigl ( \partial_\rho \overline{C}^\rho \bigr ) \bigl ( \Gamma_{\sigma \mu \nu} C^\sigma \bigr ) - 2 \bigl ( \partial_\mu \overline{C}^\rho \bigr ) \bigl ( \Gamma_{\sigma \rho \nu} C^\sigma \bigr ) \right ) \dif V_\eta \\
		& \phantom{\coloneq} + \frac{1}{4} \eta^{\mu \nu} \left ( \bigl ( \Gamma_{\sigma \mu \nu} \overline{C}^\sigma \bigr ) \bigl ( \partial_\rho C^\rho \bigr ) - 2 \bigl ( \Gamma_{\sigma \rho \nu} \overline{C}^\sigma \bigr ) \bigl ( \partial_\mu C^\rho \bigr ) \right ) \dif V_\eta \\
		& \phantom{\coloneq} + \frac{\gcoupling^2 \zeta}{8} \eta_{\mu \nu} \left ( \overline{C}^\rho \bigl ( \partial_\rho \overline{C}^\mu \bigr ) \right ) \left ( C^\sigma \bigl ( \partial_\sigma C^\nu \bigr ) \right ) \dif V_\eta \, . \label{eqn:sym-gf-ghost-gr}
	\end{split}
\end{align}
\end{subequations}
In particular, we consider the metric expansion \(g_{\mu \nu} \equiv \eta_{\mu \nu} + \gcoupling h_{\mu \nu}\), where \(h_{\mu \nu}\) is the graviton field and \(\gcoupling \coloneq \sqrt{\kappa}\) the graviton coupling constant (with \(\kappa \coloneq 8 \pi G\) the Einstein gravitational constant). In addition, \(R \coloneq g^{\nu \sigma} \tensor{R}{^\mu _\nu _\mu _\sigma}\) is the Ricci scalar with \(\tensor{R}{^\rho _\sigma _\mu _\nu} \coloneq \partial_\mu \tensor{\Gamma}{^\rho _\nu _\sigma} - \partial_\nu \tensor{\Gamma}{^\rho _\mu _\sigma} + \tensor{\Gamma}{^\rho _\mu _\lambda} \tensor{\Gamma}{^\lambda _\nu _\sigma} - \tensor{\Gamma}{^\rho _\nu _\lambda} \tensor{\Gamma}{^\lambda _\mu _\sigma}\) the Riemann tensor and \(\tensor{\Gamma}{^\rho _\mu _\nu} \coloneq g^{\rho \sigma} \textfrac{\left ( \partial_\mu g_{\sigma \nu} + \partial_\nu g_{\mu \sigma} - \partial_\sigma g_{\mu \nu} \right )}{2}\) the Christoffel symbol for the Levi-Civita connection. Furthermore, \(\dif V_\eta \coloneq \dif t \wedge \dif x \wedge \dif y \wedge \dif z\) denotes the Minkowskian volume form, which is related to the Riemannian volume form \(\dif V_g\) via \(\dif V_g \equiv \sqrt{- \dt{g}} \dif V_\eta\). Moreover, \(\lindeDonder_\mu \coloneq \eta^{\rho \sigma} \Gamma_{\mu \rho \sigma} \equiv 0\) is the linearized de Donder gauge fixing functional with \(\Gamma_{\mu \rho \sigma} \coloneq \textfrac{\left ( \partial_\rho g_{\mu \sigma} + \partial_\sigma g_{\rho \mu} - \partial_\mu g_{\rho \sigma} \right )}{2}\) and \(\zeta\) denotes the gauge fixing parameter. Finally, \(C_\mu\) and \(\overline{C}^\mu\) are the graviton-ghost and graviton-antighost, respectively. We refer to \conref{con:sign_choices} and \cite{Prinz_2,Prinz_4} for more detailed introductions and further comments on the chosen conventions.

In addition, the setup for Quantum Yang--Mills theory is given as follows:
\begin{subequations}
\begin{align}
	\mathcal{L}_\text{QYM} & \coloneq \mathcal{L}_\text{YM} + \mathcal{L}_\text{YM-GF-Ghost}
	\intertext{with the Yang--Mills Lagrange density}
	\mathcal{L}_\text{YM} & \coloneq - \frac{1}{4 \mathrm{g}^2} \delta_{a b} g^{\mu \nu} g^{\rho \sigma} F^a_{\mu \rho} F^b_{\nu \sigma} \dif V_g
	\intertext{and the symmetric gauge fixing and ghost Lagrange density, in the form derived in \cite{Prinz_6}, cf.\ \cite{Baulieu_Thierry-Mieg},}
	\begin{split}
		\mathcal{L}_\text{YM-GF-Ghost} & \coloneq \frac{1}{\xi} \left ( - \frac{1}{2 \mathrm{g}^2} \delta_{ab} \Lorenz^a \Lorenz^b + g^{\mu \nu} \left ( \partial_\mu \overline{c}_a \right ) \left ( \partial_\nu c^a \right ) \right ) \dif V_g \\
		& \phantom{\coloneq} + \frac{\mathrm{g}}{2} g^{\mu \nu} \tensor{f}{^a _b _c} \left ( \bigl ( \partial_\mu \overline{c}_a \bigr ) \bigl ( c^b A^c_\nu \bigr ) + \bigl ( \overline{c}_a A^b_\nu \bigr ) \bigl ( \partial_\mu c^c \bigr ) \right ) \dif V_g \\
		& \phantom{\coloneq} + \frac{\mathrm{g}^2 \xi}{16} f^{a b c} f_{ade} \overline{c}_b \overline{c}_c c^d c^e \dif V_g \, . \label{eqn:sym-gf-ghost-ym}
	\end{split}
\end{align}
\end{subequations}
We remark that \(F^a_{\mu \nu} \coloneq \mathrm{g} \bigl ( \partial_\mu A^a_\nu - \partial_\nu A^a_\mu \bigr ) - \mathrm{g}^2 \tensor{f}{^a _b _c} A^b_\mu A^c_\nu\) is the local curvature form of the gauge boson \(A^a_\mu\). Furthermore, \(\dif V_g \coloneq \sqrt{- \dt{g}} \dif t \wedge \dif x \wedge \dif y \wedge \dif z\) denotes again the Riemannian volume form. In addition, \(\Lorenz^a \coloneq \mathrm{g} g^{\mu \nu} \bigl ( \nabla^{TM}_\mu A^a_\nu \bigr ) \equiv 0\) denotes the covariant Lorenz gauge fixing functional and \(\xi\) is the gauge fixing parameter. Finally, \(c^a\) and \(\overline{c}_a\) are the gauge ghost and gauge antighost, respectively.

Finally, the setup for matter in the Standard Model is represented via a vector of complex scalar fields and a vector of spinor fields, both subjected to the action of the gauge group. Specifically, they are given as follows:
\begin{subequations} \label{eqn:matter_lagrange_density_introduction}
\begin{align}
	\mathcal{L}_\text{Matter} & \coloneq \mathcal{L}_\text{Higgs} + \mathcal{L}_\text{Fermion}
	\intertext{with the Higgs Lagrange density}
	\mathcal{L}_\text{Higgs} & \coloneq \left ( g^{\mu \nu} \left ( \nabla^{H}_\mu \Phi \right )^\dagger \left ( \nabla^{H}_\nu \Phi \right ) + \sum_{i \in \boldsymbol{I}_\Phi} \frac{\alpha_i}{i!} \bigl ( \Phi^\dagger \Phi \bigr )^i \right ) \dif V_g
	\intertext{and the symmetrized fermion Lagrange density}
	\mathcal{L}_\text{Fermion} & \coloneq \left ( \frac{\imaginary}{2} \overline{\Psi} \left ( \slashed{\nabla}^{\boldsymbol{\Sigma} M} \Psi \right ) - \frac{\imaginary}{2} \overline{\left ( \slashed{\nabla}^{\boldsymbol{\Sigma} M} \Psi \right )} \Psi - \boldsymbol{m}_\Psi \overline{\Psi} \Psi \right ) \dif V_g \, . \label{eqn:sym-fermion}
\end{align}
\end{subequations}
Here, \(\Phi\) and \(\Psi\) denote the respective vectors of complex scalar fields and spinor fields, with corresponding dual vectors \(\Phi^\dagger\) and dual spinors \(\overline{\Psi} \coloneq \left ( \boldsymbol{\gamma}_0 \Psi \right )^\dagger\). Furthermore, \(\nabla^{H}_\mu \coloneq \partial_\mu + \imaginary \mathrm{g} A^a_\mu \mathfrak{H}_a\) and \(\nabla^{\boldsymbol{\Sigma} M}_\mu \coloneq \partial_\mu + \boldsymbol{\varpi}_\mu + \imaginary \mathrm{g} A^a_\mu \mathfrak{S}_a\) denote the respective covariant derivatives, where \(\mathfrak{H}_a\) and \(\mathfrak{S}_a\) denote the infinitesimal actions of the gauge group \(G\) on the Higgs bundle \(H\) and the twisted spinor bundle \(\boldsymbol{\Sigma} M\), respectively, and \(\boldsymbol{\varpi}_\mu\) is the spin connection on the twisted spinor bundle. In addition, \(\slashed{\nabla}^{\boldsymbol{\Sigma} M} \coloneq e^{\mu m} \gamma_m \bigl ( \partial_\mu + \boldsymbol{\varpi}_\mu + \imaginary \mathrm{g} A^a_\mu \mathfrak{S}_a \bigr )\) denotes the corresponding twisted Dirac operator, where \(e^{\mu m}\) is the inverse vielbein and \(\gamma_m\) the Minkowski space Dirac matrix. Moreover, \(\boldsymbol{I}_\Phi\) denotes the set of scalar field interactions, with respective coupling constants \(\alpha_i\) (and possible mass \(\alpha_1 \coloneq - m_\Phi^2\)). Finally, \(\boldsymbol{m}_\Psi\) denotes the diagonal matrix with all fermion masses as entries. We refer to \cite[Subsection 4.2]{Prinz_4} for a detailed discussion thereon.

This article is organized as follows: We start by displaying the QGR-SM Feynman rules for all propagators and all three-valent vertices in \sectionref{sec:explicit_feynman_rules}. Then we proceed by studying their longitudinal and transversal properties in \sectionref{sec:longitudinal_and_transversal_projections}. More precisely, we start in \ssecref{ssec:transversality_qym} by recalling known and immediate identities in QYM. Then we proceed in \ssecref{ssec:transversality_qgr} by introducing their novel and involved counterparts in QGR. Specifically, our results are as follows: We introduce the notion of an \emph{optimal gauge fixing} in \defnref{defn:optimal-gauge-fixing}: This is a gauge fixing that, for a given gauge theory, acts only on the vertical (i.e.\ gauge) degrees of freedom in a covariant way. Thus, it complements the Lagrange density of the gauge theory in a unique way. In particular, we show that the Lorenz gauge fixing for Yang--Mills theory and the de Donder gauge fixing for General Relativity are both optimal in \thmsaref{thm:feynman-rule-gluon-propagator-lt-decomposition}{thm:feynman-rule-graviton-propagator-lt-decomposition}, which highlights their special roles. Then we present the respective \emph{transversal structures}, that is the sets of longitudinal, identical and transversal projection tensors introduced in \defnref{defn:qgr-transversal-structure}, together with their corresponding metrics: First we recall the situation of QYM in \defnref{defn:qym-transversal-structure}, which is given by
\begin{equation}
	\mathcal{T}_\text{QYM} \coloneq \set{L, I, T}_\mettransstr
\end{equation}
with
\begin{subequations}
\begin{align}
	L^\nu_\mu & \coloneq \frac{1}{p^2} p^\nu p_\mu \, , \\
	I^\nu_\mu & \coloneq \delta^\nu_\mu \, , \vphantom{\frac{1}{p^2}} \\
	T^\nu_\mu & \coloneq I^\nu_\mu - L^\nu_\mu \vphantom{\frac{1}{p^2}}
\end{align}
\end{subequations}
and metric
\begin{subequations}
\begin{align}
	\mettransstr_{\mu \nu} & \coloneq \frac{1}{p^2} \eta_{\mu \nu} \label{eqn:gluon-trans-metric-introduction}
	\intertext{with inverse}
	\mettransstr^{\mu \nu} & \coloneq p^2 \eta^{\mu \nu} \, , \label{eqn:gluon-trans-invmetric-introduction}
\end{align}
\end{subequations}
satisfying \(\mettransstr_{\mu \tau} \mettransstr^{\tau \nu} = I^\nu_\mu\) and the following relations, cf.\ \thmref{thm:qym-transversal-structure}:
\begin{subequations}
\begin{align}
	& L_\mu^\tau L_\tau^\nu = L_\mu^\nu \, , \quad && I_\mu^\tau I_\tau^\nu = I_\mu^\nu \, , \quad & T_\mu^\tau T_\tau^\nu = T_\mu^\nu \, , \\
	& L_\mu^\tau I_\tau^\nu = L_\mu^\nu \, , \quad && T_\mu^\tau I_\tau^\nu = T_\mu^\nu \, , \quad & L_\mu^\tau T_\tau^\nu = 0 \, .
\end{align}
\end{subequations}
Then, we introduce the corresponding counterpart of QGR in \defnref{defn:qgr-transversal-structure}, which is given by
\begin{equation}
	\mathcal{T}_\text{QGR} \coloneq \set{\mathbbit{L}, \mathbbit{I}, \mathbbit{T}}_{\mathbbit{\mettransstr}}
\end{equation}
with
\begin{subequations}
\begin{align}
	\mathbbit{L}^{\rho \sigma}_{\mu \nu} & \coloneq \frac{1}{2 p^2} \left ( \delta^\rho_\mu p^\sigma p_\nu + \delta^\sigma_\mu p^\rho p_\nu + \delta^\rho_\nu p^\sigma p_\mu + \delta^\sigma_\nu p^\rho p_\mu - 2 \eta^{\rho \sigma} p_\mu p_\nu \right ) \, , \\
	\mathbbit{I} \mspace{2mu} ^{\rho \sigma}_{\mu \nu} & \coloneq \frac{1}{2} \left ( \delta^\rho_\mu \delta^\sigma_\nu + \delta^\sigma_\mu \delta^\rho_\nu \right ) \, , \vphantom{\frac{1}{p^2}} \\
	\mathbbit{T} \mspace{2mu} ^{\rho \sigma}_{\mu \nu} & \coloneq \mathbbit{I} \mspace{2mu} ^{\rho \sigma}_{\mu \nu} - \mathbbit{L}^{\rho \sigma}_{\mu \nu} \vphantom{\frac{1}{p^2}}
\end{align}
\end{subequations}
and metric\footnote{For a general spacetime dimension \(d\), \eqnref{eqn:grav-trans-metric-introduction} actually reads \begin{equation*} \mathbbit{\mettransstr}_{\mu \nu \rho \sigma} \coloneq \frac{1}{p^2} \left ( \eta_{\mu \rho} \eta_{\nu \sigma} + \eta_{\mu \sigma} \eta_{\nu \rho} - \frac{2}{d - 2} \eta_{\mu \nu} \eta_{\rho \sigma} \right )\end{equation*} with \eqnref{eqn:grav-trans-invmetric-introduction} unchanged.}
\begin{subequations}
\begin{align}
	\mathbbit{\mettransstr}_{\mu \nu \rho \sigma} & \coloneq \frac{1}{p^2} \left ( \eta_{\mu \rho} \eta_{\nu \sigma} + \eta_{\mu \sigma} \eta_{\nu \rho} - \eta_{\mu \nu} \eta_{\rho \sigma} \right ) \label{eqn:grav-trans-metric-introduction}
	\intertext{with inverse}
	\mathbbit{\mettransstr}^{\mu \nu \rho \sigma} & \coloneq \frac{p^2}{4} \left ( \eta^{\mu \rho} \eta^{\nu \sigma} + \eta^{\mu \sigma} \eta^{\nu \rho} - \eta^{\mu \nu} \eta^{\rho \sigma} \right ) \, , \label{eqn:grav-trans-invmetric-introduction}
\end{align}
\end{subequations}
satisfying \(\mathbbit{\mettransstr}_{\mu \nu \kappa \lambda} \mathbbit{\mettransstr}^{\kappa \lambda \rho \sigma} = \mathbbit{I} \mspace{1mu} ^{\rho \sigma}_{\mu \nu}\) and the following relations, cf.\ \thmref{thm:qgr-transversal-structure}:
\begin{subequations}
\begin{align}
	& \mathbbit{L}^{\kappa \lambda}_{\mu \nu} \, \mathbbit{L}^{\rho \sigma}_{\kappa \lambda} = \mathbbit{L}^{\rho \sigma}_{\mu \nu} \, , \quad && \mathbbit{I} \mspace{1mu} ^{\kappa \lambda}_{\mu \nu} \, \mathbbit{I} \mspace{1mu} ^{\rho \sigma}_{\kappa \lambda} = \mathbbit{I} \mspace{1mu} ^{\rho \sigma}_{\mu \nu} \, , \quad & \mathbbit{T}^{\kappa \lambda}_{\mu \nu} \, \mathbbit{T}^{\rho \sigma}_{\kappa \lambda} = \mathbbit{T}^{\rho \sigma}_{\mu \nu} \, , \\
	& \mathbbit{L}^{\kappa \lambda}_{\mu \nu} \, \mathbbit{I} \mspace{1mu} ^{\rho \sigma}_{\kappa \lambda} = \mathbbit{L}^{\rho \sigma}_{\mu \nu} \, , \quad && \mathbbit{T}^{\kappa \lambda}_{\mu \nu} \, \mathbbit{I} \mspace{1mu} ^{\rho \sigma}_{\kappa \lambda} = \mathbbit{T}^{\rho \sigma}_{\mu \nu} \, , \quad & \mathbbit{L}^{\kappa \lambda}_{\mu \nu} \, \mathbbit{T}^{\rho \sigma}_{\kappa \lambda} = 0 \, .
\end{align}
\end{subequations}
Thus, the complete transversal structure of QGR-SM is given via the disjoint union
\begin{equation}
	\mathcal{T}_\text{QGR-SM} \coloneq \mathcal{T}_\text{QGR} \cup \mathcal{T}_\text{QYM} \, ,
\end{equation}
with the two transversal structures mutually commuting, and acting on gravitons, gluons, photons, \(Z\)-bosons and \(W^\pm\)-bosons, respectively. Next we study the decomposition of the longitudinal projection tensor into the product of a gauge transformation together with the gauge fixing projection in \lemsaref{lem:g_and_f_inverse_decomposition_gl_qym}{lem:g_and_f_inverse_decomposition_gl_qgr}. In particular, we show that the provided longitudinal, identical and transversal projection tensors are indeed projectors in \thmsaref{thm:qym-transversal-structure}{thm:qgr-transversal-structure}. Furthermore, we show that gauge transformations and the gauge fixing projections are eigentensors of the respective transversal structures in \propsaref{prop:gf-eigenvectors-lit-qym}{prop:gf-eigentensors-lit-qgr}. Thereafter, we study the action of the corresponding metrics on said tensors in \lemsaref{lem:identities_tensors_qym}{lem:identities_tensors_qgr} and \colsaref{col:l-tensor-gg-ff-qym}{col:l-tensor-gg-ff-qgr}. This allows us ultimately to simplify the gauge boson and graviton propagators as follows:\footnote{Observe the appearing factors of \(\textfrac{1}{p^2}\) when lowering the indices on the transversal structure with the metrics, i.e.\ \eqnsaref{eqn:gluon-trans-metric-introduction}{eqn:grav-trans-metric-introduction}, to conveniently match the Feynman rules, given in \eqnsaref{eqn:fr-p-gluon}{eqn:fr-p-graviton}.}
\begin{align}
	\begin{split}
		\Phi \left ( \cgreen{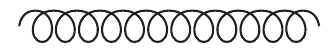} \right ) & = - \frac{\imaginary p^2}{p^2 + \imaginary \epsilon} \delta^{a b} \left ( \mettransstr_{\mu \nu} - \left ( 1 - \xi \right ) L_{\mu \nu} \right ) \\
		& = - \frac{\imaginary p^2}{p^2 + \imaginary \epsilon} \delta^{a b} \left ( T_{\mu \nu} + \xi L_{\mu \nu} \right )
	\end{split}
	\intertext{and}
		\begin{split}
		\Phi \left ( \cgreen{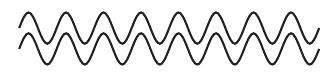} \right ) & = - \frac{2 \imaginary p^2}{p^2 + \imaginary \epsilon} \left ( \mathbbit{\mettransstr}_{\mu \nu \rho \sigma} - \left ( 1 - \zeta \right ) \mathbbit{L}_{\mu \nu \rho \sigma} \right ) \\
		& = - \frac{2 \imaginary p^2}{p^2 + \imaginary \epsilon} \left ( \mathbbit{T}_{\mu \nu \rho \sigma} + \zeta \mathbbit{L}_{\mu \nu \rho \sigma} \right ) \, .
	\end{split}
	\intertext{Furthermore, their corresponding ghost propagators relate to them via a gauge fixing projection (denoted via \(\gfym\) and \(\gfgr\), respectively) and an infinitesimal gauge transformation (denoted via \(\gtym\) and \(\gtgr\), respectively --- with \(\gtym^\prime\) and \(\gtgr^\prime\) being their prefactor reduced variants, to avoid additional factors of \(p^2\)) with the arrows indicating in which direction they act, cf.\ \defnsaref{defn:qym-transversal-structure}{defn:qgr-transversal-structure}:}
	\Phi \bigl ( {\scriptstyle \vec{\gfym}} \cgreen{p-gluon} \bigr ) & = \Phi \bigl ( \cgreen{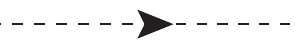} {\scriptstyle \vec{\gtym}^\prime} \bigr )
	\intertext{and}
	\Phi \bigl ( {\scriptstyle \vec{\gfgr}} \cgreen{p-graviton} \bigr ) & = \Phi \bigl ( \cgreen{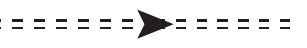} {\scriptstyle \vec{\gtgr}^\prime} \, \bigr ) \, .
\end{align}
These results can be found in \thmsaref{thm:feynman-rule-gluon-propagator-lt-decomposition}{thm:feynman-rule-graviton-propagator-lt-decomposition}, respectively. Additionally, we provide cancellation identities for the gauge boson and graviton vertex Feynman rules in \thmsaref{thm:three-valent-contraction-identities-qym}{thm:three-valent-contraction-identities-qgr}, as well as for their corresponding couplings to matter from the Standard Model (SM) in \thmsaref{thm:three-valent-contraction-identities-qym-with-matter}{thm:three-valent-contraction-identities-qgr-with-matter} and \colref{col:three-valent-contraction-identities-gravitongluontriple}. These identities have important consequences for the renormalization of QGR-SM: We have extensively discussed in \cite[Subsection 3.3]{Prinz_2} possible solutions to a fundamental problem that frequently occurs in the renormalization of gauge theories: It may be that there exist divergent amplitudes without a corresponding monomial in the Lagrange density to absorb the appearing divergences. This occurs famously in Quantum Electrodynamics with the three- and four-valent photon-interactions. Here, it turns out that the sum over all three-valent photon-interaction Feynman integrals adds up to zero by virtue of Furry's theorem.\footnote{We note the generalization of Furry's theorem to the coupling of (effective) Quantum General Relativity to Quantum Electrodynamics in \cite[Theorem 3.55]{Prinz_2}.} Additionally, it turns out that the sum over all four-valent photon-interaction Feynman integrals is finite, despite the \emph{superficial degree of divergence} suggesting otherwise \cite{Aldins_Brodsky_Dufner_Kinoshita}. We remark that similar issues of superficially divergent Feynman integrals adding up to a finite contribution also arise in other sectors of the Standard Model, notably the gluon-ghost sector: This is explained by the gauge invariant regularization procedure of 't Hooft and Veltman \cite{tHooft_Veltman_Nobel_Prize} in combination with the transversality constraint of the corresponding form factor decompositions. Crucially, there appear infinitely many such problems in the coupling of (effective) Quantum General Relativity to the Standard Model, which lack a convincing understanding so far: In particular, Feynman integrals with external matter particles and sufficiently many virtual gravitons are superficially divergent for arbitrary valences. To this end, we suggest the application of \cite[Solution 3.39]{Prinz_2}: Here, we propose to absorb the divergences of Feynman integrals into corresponding symmetrized sums over trees, with gravitons as virtual particles. In particular, with reference to \thmsaref{thm:three-valent-contraction-identities-qym-with-matter}{thm:three-valent-contraction-identities-qgr-with-matter} and \colref{col:three-valent-contraction-identities-gravitongluontriple}, we argue that this a priori non-local operation can become effectively local if corresponding Slavnov--Taylor-like identities hold (with the corresponding four-valent vertex being zero).\footnote{We remark that a similar situation also appears for gauge ghosts with the Faddeev--Popov ghost construction.} Finally, we conclude our investigations with a comment on the differences of the two most prominent definitions of the graviton field: The metric decomposition and the metric density decomposition of Goldberg \cite{Goldberg} in \remref{rem:md_vs_mdd}.

Additionally, we provide the derived QGR-SM Feynman rules also in FORM code \cite{Vermaseren,Kuipers_Ueda_Vermaseren_Vollinga,Davies_Kaneko_Marinissen_Ueda_Vermaseren} in \appendixref{sec:explicit_feynman_rules_form} and via the supplementary file \texttt{QGR-SM-FR.frm}, available on the journal homepage and the arXiv preprint server.

The present article integrates as follows into the literature: Ultimately, we aim to prove the renormalizability of (effective) Quantum General Relativity via generalized Slavnov--Taylor identities in the sense of \cite{Prinz_3,Kreimer_QG1,Kreimer_QG2}. More precisely, we aim to show them graphically via so-called cancellation identities, cf.\ \cite{Cvitanovic,tHooft_Veltman,Sars_PhD,Kissler_Kreimer,Gracey_Kissler_Kreimer,Kissler}. These identities will then be implemented on the algebra of Feynman graphs via a modified version of the Feynman graph cohomology introduced in \cite{Kreimer_Sars_vSuijlekom,Berghoff_Knispel}. In particular, we argue that the compatibility of cancellation identities with renormalization is reflected in the well-definedness of the corresponding differential-graded renormalization Hopf algebra, which will be introduced in \cite{Prinz_9}. Additionally, we argue there that this differential-graded renormalization Hopf algebra constitutes a perturbative version of BRST cohomology \cite{Becchi_Rouet_Stora_1,Becchi_Rouet_Stora_2,Becchi_Rouet_Stora_3,Tyutin}: BRST cohomology is a powerful tool to study gauge fixing and ghost Lagrange densities of gauge theories. Furthermore, we highlight the interesting connection between Feynman graph cohomology and the Corolla polynomial, which constructs a relation between \(\phi^3_4\) scalar field theory amplitudes and Quantum Yang--Mills theory amplitudes \cite{Kreimer_Sars_vSuijlekom,Kreimer_Yeats,Prinz_1,Kreimer_Corolla}.

This article originates from the author's dissertation \cite{Prinz_PhD}.

\section{Explicit Feynman rules} \label{sec:explicit_feynman_rules}

We start this article by providing the concrete gravity-matter Feynman rules for all propagators and three-valent vertices of (effective) Quantum General Relativity coupled to the Standard Model (QGR-SM). We emphasize again that in this article we use the symmetrized fermion Lagrange density, given in \eqnref{eqn:sym-fermion}, and the symmetric (hermitian) ghost Lagrange densities associated to the Lorenz and de Donder gauge fixing conditions, respectively, given in \eqnsaref{eqn:sym-gf-ghost-gr}{eqn:sym-gf-ghost-ym}. The symmetrized fermion Lagrange density originates from the more common asymmetric one via a partial integration with respect to the spin connection. In addition, the symmetric (hermitian) graviton-ghost and gauge ghost Lagrange densities are properly derived in \cite{Prinz_6}: This is achieved using the BRST double complex of diffeomorphisms and gauge transformations, introduced in \cite{Prinz_5}, to generalize the results of \cite{Baulieu_Thierry-Mieg} to covariant Quantum Yang--Mills theory and (effective) Quantum General Relativity. We note the presence of four-valent gauge ghost and graviton-ghost interactions in the symmetric setup and highlight the benefit of significantly simpler cancellation identities, which is the reason for choosing them here, cf.\ \thmsaref{thm:three-valent-contraction-identities-qym-with-matter}{thm:three-valent-contraction-identities-qgr-with-matter} and \colref{col:three-valent-contraction-identities-gravitongluontriple}. Additionally, we acknowledge the use of \texttt{JaxoDraw} for drawing the propagator and vertex Feynman graphs \cite{Binosi_Theussl,Binosi_Collins_Kaufhold_Theussl}. Finally, we also list the Feynman rules below in \appendixref{sec:explicit_feynman_rules_form} in FORM code and the supplementary file \texttt{QGR-SM-FR.frm}, which is available on the journal homepage and the arXiv preprint server, so that they can be readily used in computer calculations. For concreteness, we use the following sign conventions:

\enter

\begin{con}[Sign choices] \label{con:sign_choices}
	We use the sign-convention \((-++)\), as classified by \cite{Misner_Thorne_Wheeler}, i.e.:
	\begin{enumerate}
		\item Minkowski metric: \(\eta_{\mu \nu} = \begin{pmatrix} 1 & 0 & 0 & 0 \\ 0 & - 1 & 0 & 0 \\ 0 & 0 & - 1 & 0 \\ 0 & 0 & 0 & - 1 \end{pmatrix}_{\mu \nu}\)
		\item Riemann tensor: \(\tensor{R}{^\rho _\sigma _\mu _\nu} = \partial_\mu \Gamma^\rho_{\nu \sigma} - \partial_\nu \Gamma^\rho_{\mu \sigma} + \Gamma^\rho_{\mu \lambda} \Gamma^\lambda_{\nu \sigma} - \Gamma^\rho_{\nu \lambda} \Gamma^\lambda_{\mu \sigma}\)
		\item Einstein field equations: \(G_{\mu \nu} = \kappa T_{\mu \nu}\)
	\end{enumerate}
	Furthermore, we use the plus-signed Clifford relation, i.e.\ \(\left \{ \gamma_\mu , \gamma_\nu \right \} = 2 g_{\mu \nu} \id_{\Sigma \mathbbit{M}}\). Moreover, the used sign conventions for the gauge fixing and ghost constructions underlying the stated Lagrange densities are established in \cite{Prinz_5,Prinz_6}. We remark that these choices are consistent throughout the author's works, notably \cite{Prinz_2,Prinz_4,Prinz_8,Prinz_PhD}.
\end{con}

\subsection{QGR-SM propagators} \label{ssec:gravity-matter-propagators}

The arrows denote the particle flow (if applicable); all momenta are considered from left to right. We note that our graviton propagator Feynman rule in \eqnref{eqn:fr-p-graviton} differs from the main literature, e.g.\ \cite{Choi_Shim_Song}, in that we do not assume the Feynman gauge \(\zeta \coloneq 1\) for the linearized de Donder gauge fixing condition \(\lindeDonder_\mu \coloneq \eta^{\rho \sigma} \Gamma_{\mu \rho \sigma}\) and instead consider a general gauge parameter \(\zeta \in \mathbb{R}\), cf.\ \cite[Theorem 4.12]{Prinz_4}. Additionally, unlike e.g.\ \cite{Romao_Silva}, we use rescaled ghost Lagrange densities such that the gauge ghost and graviton-ghost propagators \eqnsaref{eqn:fr-p-gluonghost}{eqn:fr-p-gravitonghost} have the same scaling in \(\xi\) and \(\zeta\) as the longitudinal contributions of the corresponding graviton and gauge boson propagators \eqnsaref{eqn:fr-p-gluon}{eqn:fr-p-graviton}, cf.\ \cite{Prinz_5,Prinz_6} for the corresponding BRST setup:
{\allowdisplaybreaks
\begin{align}
	\Phi \left ( \scriptstyle i \cgreen{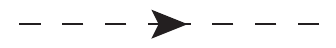} j \right ) & = \frac{\imaginary}{p^2 - m^2 + \imaginary \epsilon} \delta^{i j} \label{eqn:fr-p-scalar} \\
	\Phi \left ( \scriptstyle k \cgreen{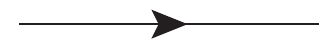} l \right ) & = \frac{\imaginary \bigl ( \slashed{p} + m \bigr )}{p^2 - m^2 + \imaginary \epsilon} \delta^{k l} \label{eqn:fr-p-spinor} \\
	\Phi \left ( \scriptstyle \mu , a \cgreen{p-gluon} \nu , b \right ) & = - \frac{\imaginary}{p^2 + \imaginary \epsilon} \delta^{a b} \left ( \eta_{\mu \nu} - \frac{\left ( 1 - \xi \right )}{p^2} p_\mu p_\nu \right ) \label{eqn:fr-p-gluon} \\
	\Phi \left ( \scriptstyle a \cgreen{p-gluonghost} b \right ) & = - \frac{\imaginary \xi}{p^2 + \imaginary \epsilon} \delta^{a b} \label{eqn:fr-p-gluonghost} \\
	\begin{split} \label{eqn:fr-p-graviton}
		\Phi \left ( \scriptstyle \mu \nu \cgreen{p-graviton} \rho \sigma \right ) & = - \frac{2 \imaginary}{p^2 + \imaginary \epsilon} \Biggl ( \left ( \eta_{\mu \rho} \eta_{\nu \sigma} + \eta_{\mu \sigma} \eta_{\nu \rho} - \eta_{\mu \nu} \eta_{\rho \sigma} \right ) \\
		& \phantom{= - [} - \left ( \frac{1 - \zeta}{p^2} \right ) \left ( \eta_{\mu \rho} p_\nu p_\sigma + \eta_{\mu \sigma} p_\nu p_\rho + \eta_{\nu \rho} p_\mu p_\sigma + \eta_{\nu \sigma} p_\mu p_\rho \right ) \Biggr )
	\end{split} \\
	\Phi \left ( \scriptstyle \rho \cgreen{p-gravitonghost} \sigma \right ) & = - \frac{2 \imaginary \zeta}{p^2 + \imaginary \epsilon} \eta_{\rho \sigma} \label{eqn:fr-p-gravitonghost}
\end{align}
}%

\subsection{QGR-SM 3-valent vertices} \label{ssec:three-val-gravity-matter-vertices}

The arrows denote the particle flow (if applicable); all momenta are considered incoming. In addition, the sum over \(\operatorname{Perm} \left ( 1, 2, 3 \right )\) runs over all permutations of \(\set{1, 2, 3}\) and the sum over \(\mu_i \leftrightarrow \nu_i\) means symmetrization between \(\mu_i\) and \(\nu_i\). We remark that our graviton vertex Feynman rule in \eqnref{eqn:fr-v-gravitontriple} differs from the main literature, e.g.\ \cite{Choi_Shim_Song}, in that we do not assume the de Donder relation \(\partial^\nu h_{\mu \nu} = \eta^{\rho \sigma} \partial_\mu h_{\rho \sigma}\) to simplify its derivation, cf.\ \cite[Theorem 4.10]{Prinz_4}. Also, we present the graviton vertex Feynman rule in a condensed expression, while the fully symmetrized expression can be found in the \texttt{QGR-SM-FR.frm} supplementary file, available on the journal homepage and the arXiv preprint server. Additionally, unlike e.g.\ \cite{Romao_Silva}, we also use a general gauge parameter \(\xi \in \mathbb{R}\) for the graviton--gauge-boson vertex in \eqnref{eqn:fr-v-gravitongluontriple} and our rescaling convention for the ghost Lagrange densities also affects the graviton--gluon-ghost vertex Feynman rule in \eqnref{eqn:fr-v-gravitongluonghosttriple}, cf.\ again \cite{Prinz_5,Prinz_6} for the corresponding BRST setup:
{\allowdisplaybreaks
\begin{align}
	\Phi \left ( \scriptstyle \rho, a \tcgreen{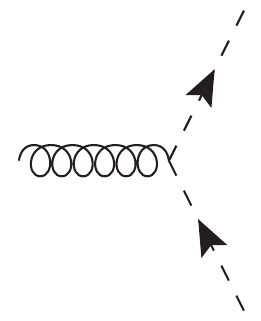}^{\scriptstyle q_2, j}_{\scriptstyle q_1, i} \right ) & = - \mathrm{g} \left ( q_1 - q_2 \right )^\rho \left ( \mathfrak{H}_a \right )_{i j} \label{eqn:fr-v-gluonscalartriple} \\
	\Phi \left ( \scriptstyle \rho, a \tcgreen{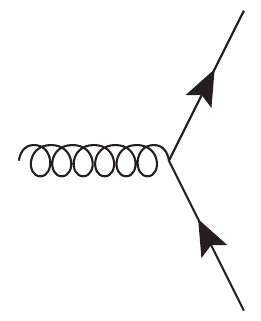}^{\scriptstyle l}_{\scriptstyle k} \right ) & = - \imaginary \mathrm{g} \gamma^\rho \left ( \mathfrak{S}_a \right )_{k l} \label{eqn:fr-v-gluonspinortriple} \\
	\Phi \left ( \scriptstyle 1 \tcgreen{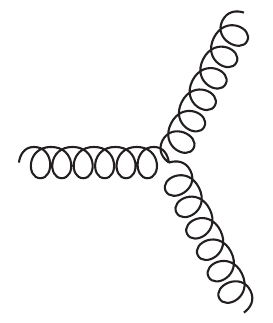}^{\scriptstyle 3}_{\scriptstyle 2} \rule{0pt}{5ex} \right ) & = - \mathrm{g} f_{a_1 a_2 a_3} \biggl ( \eta^{\rho_1 \rho_2} \left ( p_1 - p_2 \right )^{\rho_3} + \eta^{\rho_2 \rho_3} \left ( p_2 - p_3 \right )^{\rho_1} + \eta^{\rho_3 \rho_1} \left ( p_3 - p_1 \right )^{\rho_2} \biggr ) \label{eqn:fr-v-gluontriple} \\
	\Phi \left ( \scriptstyle \rho , a \tcgreen{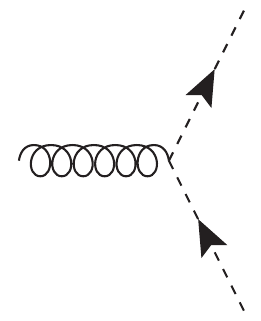}^{\scriptstyle q_2, c}_{\scriptstyle q_1, b} \right ) & = \frac{\mathrm{g}}{2} f_{a b c} \left ( q_1 - q_2 \right )^\rho \label{eqn:fr-v-gluonghosttriple} \\
	\Phi \left ( \scriptstyle \mu \nu \tcgreen{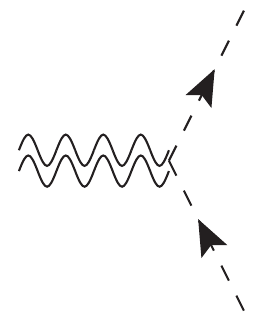}^{\scriptstyle q_2, j}_{\scriptstyle q_1, i} \right ) & = \frac{\imaginary \gcoupling}{2} \delta_{i j} \biggl ( q_1^\mu q_2^\nu + q_1^\nu q_2^\mu - \eta^{\mu \nu} \left ( q_1 \cdot q_2 + m^2 \right ) \biggr ) \label{eqn:fr-v-gravitonscalartriple} \\
	\Phi \left ( \scriptstyle \mu \nu \tcgreen{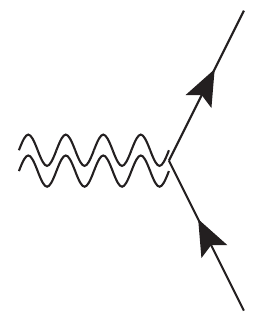}^{\scriptstyle q_2, l}_{\scriptstyle q_1, k} \right ) & = \frac{\gcoupling}{8} \delta_{k l} \biggl ( 2 \eta^{\mu \nu} \left ( \slashed{q_1} - \slashed{q_2} - 2m \right ) - \left ( q_1 - q_2 \right )^\mu \gamma^\nu - \left ( q_1 - q_2 \right )^\nu \gamma^\mu \biggr ) \label{eqn:fr-v-gravitonspinortriple} \\
	\begin{split} \label{eqn:fr-v-gravitongluontriple}
		\Phi \left ( \scriptstyle \mu \nu \tcgreen{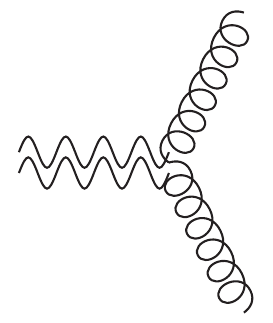}^{\scriptstyle 2}_{\scriptstyle 1} \rule{0pt}{5ex} \right ) & = \frac{\imaginary \gcoupling}{2} \delta_{a_1 a_2} \Biggl ( \left ( q_1 \cdot q_2 \right ) \left ( \eta^{\mu \nu} \eta^{\rho_1 \rho_2} - \eta^{\mu \rho_1} \eta^{\nu \rho_2} - \eta^{\mu \rho_2} \eta^{\nu \rho_1} \right ) \\
		& \phantom{= \frac{\imaginary \gcoupling}{2} \delta_{a_1 a_2} \biggl (} - \eta^{\mu \nu} q_1^{\rho_2} q_2^{\rho_1} - \eta^{\rho_1 \rho_2} \left ( q_1^\mu q_2^\nu + q_2^\mu q_1^\nu \right ) \\
		& \phantom{= \frac{\imaginary \gcoupling}{2} \delta_{a_1 a_2} \biggl (} + q_1^{\rho_2} \left ( \eta^{\mu \rho_1} q_2^\nu + \eta^{\nu \rho_1} q_2^\mu \right ) + q_2^{\rho_1} \left ( \eta^{\mu \rho_2} q_1^\nu + \eta^{\nu \rho_2} q_1^\mu \right ) \\
		& \phantom{= \frac{\imaginary \gcoupling}{2} \delta_{a_1 a_2} \biggl (} + \frac{1}{\xi} \eta^{\mu \nu} \left ( q_1^{\rho_1} q_2^{\rho_2} + q_1^{\rho_1} q_1^{\rho_2} + q_2^{\rho_1} q_2^{\rho_2} \right ) \\
		& \phantom{= \frac{\imaginary \gcoupling}{2} \delta_{a_1 a_2} \biggl (} - \frac{1}{\xi} q_1^{\rho_1} \left ( q_1^\nu \eta^{\mu \rho_2} + q_1^\mu \eta^{\nu \rho_2} \right ) - \frac{1}{\xi} q_2^{\rho_2} \left( q_2^\nu \eta^{\mu \rho_1} + q_2^\mu \eta^{\nu \rho_1} \right ) \Biggr )
	\end{split} \\
	\Phi \left ( \scriptstyle \mu \nu \tcgreen{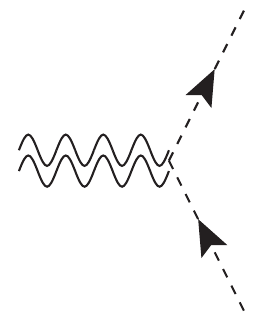}^{\scriptstyle q_2, b}_{\scriptstyle q_1, a} \right ) & = \frac{\imaginary \gcoupling}{2 \xi} \delta_{a b} \biggl ( \left ( q_1 \cdot q_2 \right ) \eta^{\mu \nu} - q_1^\mu q_2^\nu - q_2^\mu q_1^\nu \biggr ) \label{eqn:fr-v-gravitongluonghosttriple} \\
	\begin{split} \label{eqn:fr-v-gravitontriple}
		\Phi \left ( \scriptstyle 1 \tcgreen{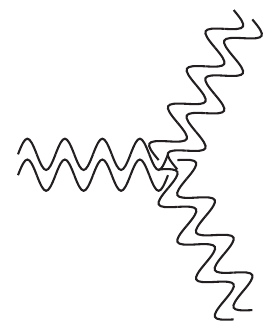}^{\scriptstyle 3}_{\scriptstyle 2} \rule{0pt}{5ex} \right ) & = - \frac{\imaginary \gcoupling}{32} \sum_{\substack{\operatorname{Perm} \left ( 1, 2, 3 \right ) \\ \mu_i \leftrightarrow \nu_i}} \\
		& \phantom{= - \frac{\imaginary \gcoupling}{32}} \Biggl ( 2 p_1^{\nu_2} p_2^{\nu_3} \eta^{\mu_1 \mu_2} \eta^{\mu_3 \nu_1} + p_1^{\nu_2} p_2^{\nu_1} \eta^{\mu_1 \mu_3} \eta^{\mu_2 \nu_3} - p_1^{\nu_1} p_2^{\nu_3} \eta^{\mu_1 \mu_3} \eta^{\mu_2 \nu_2} \\
		& \phantom{= - \frac{\imaginary \gcoupling}{32}} - p_1^{\nu_3} p_2^{\nu_1} \eta^{\mu_1 \mu_3} \eta^{\mu_2 \nu_2} - p_1^{\mu_2} p_2^{\nu_2} \eta^{\mu_1 \mu_3} \eta^{\nu_1 \nu_3} + \frac{1}{2} p_1^{\mu_3} p_2^{\nu_3} \eta^{\mu_1 \nu_1} \eta^{\mu_2 \nu_2} \\
		& \phantom{= - \frac{\imaginary \gcoupling}{32}} + \frac{1}{2} p_1^{\mu_2} p_2^{\nu_2} \eta^{\mu_1 \nu_1} \eta^{\mu_3 \nu_3} - \frac{1}{2} p_1^{\nu_2} p_2^{\nu_1} \eta^{\mu_1 \mu_2} \eta^{\mu_3 \nu_3} - \frac{1}{2} p_1^{\mu_3} p_2^{\nu_3} \eta^{\mu_1 \mu_2} \eta^{\nu_1 \nu_2} \\
		& \phantom{= - \frac{\imaginary \gcoupling}{32}} + \left ( p_1 \cdot p_2 \right ) \eta^{\mu_1 \mu_3} \eta^{\mu_2 \nu_2} \eta^{\nu_1 \nu_3} - \left ( p_1 \cdot p_2 \right ) \eta^{\mu_1 \mu_2} \eta^{\mu_3 \nu_1} \eta^{\nu_2 \nu_3} \\
		& \phantom{= - \frac{\imaginary \gcoupling}{32}} + \frac{1}{4} \left ( p_1 \cdot p_2 \right ) \eta^{\mu_1 \mu_2} \eta^{\mu_3 \nu_3} \eta^{\nu_1 \nu_2} - \frac{1}{4} \left ( p_1 \cdot p_2 \right ) \eta^{\mu_1 \nu_1} \eta^{\mu_2 \nu_2} \eta^{\mu_3 \nu_3} \Biggr )
	\end{split} \\
	\begin{split} \label{eqn:fr-v-gravitonghosttriple}
		\Phi \left ( \scriptstyle \mu \nu \tcgreen{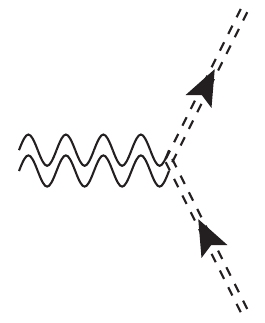}^{\scriptstyle 2}_{\scriptstyle 1} \right ) & = \frac{\imaginary \gcoupling}{8} \Biggl ( \Bigl ( q_2^\mu \bigl ( q_2^{\sigma_1} + 2 q_1^{\sigma_1} \bigr ) - q_1^\mu q_2^{\sigma_1} \Bigr ) \eta^{\nu \sigma_2} + \Bigl ( q_1^\mu \bigl ( q_1^{\sigma_2} + 2 q_2^{\sigma_2} \bigr ) - q_2^\mu q_1^{\sigma_2} \Bigr ) \eta^{\nu \sigma_1} \\
		& + \Bigl ( q_2^\nu \bigl ( q_2^{\sigma_1} + 2 q_1^{\sigma_1} \bigr ) - q_1^\nu q_2^{\sigma_1} \Bigr ) \eta^{\mu \sigma_2} + \Bigl ( q_1^\nu \bigl ( q_1^{\sigma_2} + 2 q_2^{\sigma_2} \bigr ) - q_2^\nu q_1^{\sigma_2} \Bigr ) \eta^{\mu \sigma_1} \\
		& - \bigl ( q_1^{\sigma_1} q_1^{\sigma_2} + 2 q_1^{\sigma_1} q_2^{\sigma_2} + q_2^{\sigma_1} q_2^{\sigma_2} \bigr ) \eta^{\mu \nu} - \left ( q_1 + q_2 \right )^2 \bigl ( \eta^{\mu \sigma_1} \eta^{\nu \sigma_2} + \eta^{\mu \sigma_2} \eta^{\nu \sigma_1} \bigr ) \Biggr )
	\end{split}
\end{align}
}%

\section{Longitudinal and transversal projections} \label{sec:longitudinal_and_transversal_projections}

In this section, we study the longitudinal and transversal properties of Quantum Yang--Mills theory with a Lorenz gauge fixing (QYM) and (effective) Quantum General Relativity with a de Donder gauge fixing (QGR). In particular, we start with known and immediate identities in QYM and then proceed to their novel and involved counterparts in QGR. To this end, we first introduce the notion of an \emph{optimal gauge fixing} in \defnref{defn:optimal-gauge-fixing}: This is a gauge fixing that, for a given gauge theory, acts only on the vertical (i.e.\ gauge) degrees of freedom and thus complements the Lagrange density of the gauge theory in a unique way. In particular, we show that the Lorenz gauge fixing for Quantum Yang--Mills theory and the de Donder gauge fixing for (effective) Quantum General Relativity are both optimal in \thmsaref{thm:feynman-rule-gluon-propagator-lt-decomposition}{thm:feynman-rule-graviton-propagator-lt-decomposition}, which highlights their special roles. Then we present the respective transversal structures, i.e.\ the sets of longitudinal, identical and transversal projection operators introduced in \cite[Definition 2.4]{Prinz_3}, together with their corresponding metrics: First we recall the situation of QYM in \defnref{defn:qym-transversal-structure} and then we introduce the corresponding counterpart of QGR in \defnref{defn:qgr-transversal-structure}. Then we study the decomposition of the longitudinal projection operators into the product of the respective gauge transformation and gauge fixing projections in \lemsaref{lem:g_and_f_inverse_decomposition_gl_qym}{lem:g_and_f_inverse_decomposition_gl_qgr}. In addition, we show that the provided longitudinal, identical and transversal projection operators are indeed projectors in \thmsaref{thm:qym-transversal-structure}{thm:qgr-transversal-structure}. Furthermore, we show that the gauge transformation and gauge fixing projections are eigentensors of the respective transversal structures in \propsaref{prop:gf-eigenvectors-lit-qym}{prop:gf-eigentensors-lit-qgr}. Thereafter, we study the action of the corresponding metrics on said tensors in \lemsaref{lem:identities_tensors_qym}{lem:identities_tensors_qgr} and \colsaref{col:l-tensor-gg-ff-qym}{col:l-tensor-gg-ff-qgr}. This allows us then finally to simplify the gauge boson and graviton propagators and relate them to their ghost propagators via the gauge fixing projections in \thmsaref{thm:feynman-rule-gluon-propagator-lt-decomposition}{thm:feynman-rule-graviton-propagator-lt-decomposition}. Additionally, we provide cancellation identities for the gauge boson and graviton vertex Feynman rules in \thmsaref{thm:three-valent-contraction-identities-qym}{thm:three-valent-contraction-identities-qgr}, as well as for the corresponding couplings to matter from the Standard Model in \thmsaref{thm:three-valent-contraction-identities-qym-with-matter}{thm:three-valent-contraction-identities-qgr-with-matter} and \colref{col:three-valent-contraction-identities-gravitongluontriple}. These identities then suggest the use of \cite[Solution 3.39]{Prinz_2} for gravity-matter couplings: Higher-valent matter Feynman diagrams are divergent if the virtual particles of the Feynman graphs are gravitons (e.g.\ scalar, spinor or photon four-point diagrams).\footnote{A similar situation appears for gauge ghosts with the Faddeev--Popov ghost construction.} This is problematic, as there exist no such residues in the respective Lagrange densities. However, we suggest the absorption of these divergences via corresponding trees and claim that the corresponding cancellation identities render these a priori non-local operations local. More precisely, this resembles the Slavnov--Taylor identities in Quantum Yang--Mills theory, which also relate local residues (four-valent interaction-vertex) with non-local trees (two three-valent vertices glued together with a propagator); the only difference here is the lack of a four-valent vertex in the first place. Finally, we complete our investigations with a comment on the differences of the two most prominent definitions of the graviton field: The metric decomposition and the metric density decomposition of Goldberg and Capper et al.\ in \remref{rem:md_vs_mdd}.

\enter

\begin{defn}[Transversal and physical structures] \label{defn:transversal-structure}
	We define the \emph{transversal structure} \(\TQ \coloneq \set{\boldsymbol{L}, \boldsymbol{I}, \boldsymbol{T}}_{\boldsymbol{\mettransstr}}\) of a QFT \(\Q\) as the set of the following three projection operators, acting on the Lorentz structure of propagators or vertices (half-edges in general), together with a corresponding metric to raise and lower their Lorentz indices, such that the following relations are satisfied:
	\begin{itemize}
		\item \(\boldsymbol{L} \left ( p^\mu \right )\) is the \emph{longitudinal projection operator} and related to the product of a gauge fixing projection and an infinitesimal linearized gauge transformation (see below)
		\item \(\boldsymbol{I}\) is the \emph{identity operator}
		\item \(\boldsymbol{T} \left ( p^\mu \right ) \coloneq \boldsymbol{I} - \boldsymbol{L} \left ( p^\mu \right )\) is the \emph{transversal projection operator} as the respective complement
		\item \(\boldsymbol{\mettransstr} \left ( p^\mu \right )\) denotes the metric to raise and lower the indices of \(\set{\boldsymbol{L}, \boldsymbol{I}, \boldsymbol{T}}\)
	\end{itemize}
Furthermore, we define the \emph{physical projection operator} \(\boldsymbol{P} \left ( p^\mu, m \right )\) as the composition of the above \emph{transversal projection operator} \(\boldsymbol{T} \left ( p^\mu \right )\) with the \emph{on-shell reduction operator} \(\OS \left ( p^\mu, m \right )\), as follows:
\begin{itemize}
	\item \(\OS \left ( p^\mu, m \right )\) restricts the expression to \(p^2 = m^2\) for bosons and \(\slashed{p} = m \slashed{\boldsymbol{1}}\) for fermions, where \(m\) is the mass of the corresponding particle (potentially zero)
	\item \(\boldsymbol{P} \left ( p^\mu, m \right ) \coloneq \boldsymbol{T} \left ( p^\mu \right ) \circ \OS \left ( p^\mu, m \right )\) then constitutes the \emph{physical projection operator}\footnote{We choose this order to connect to diagrammatic cancellation identities: The on-shell reduction eliminates propagator cancellations, which relate to higher-valent vertex cancellations. Then, the transversal projection eliminates the longitudinal modes, which cancel against the corresponding ghost contributions.}
\end{itemize}
In a more graph-theoretic context, we may also write \(\boldsymbol{L}_h, \boldsymbol{I}_h, \boldsymbol{T}_h, \boldsymbol{\mettransstr}_h, \OS_h, \boldsymbol{P}_h\), with \(h\) the corresponding half-edge, to which we would like to apply the operators. Finally, if the context is clear (i.e.\ when attaching them to half-edges of a Feynman graph), we may drop the dependence altogether. Explicitly, we denote the projection operators for Yang--Mills theory via \(\mathcal{T}_\text{QYM} \coloneq \set{L, I, T}_\mettransstr\) with \emph{physical projection operator} \(P\) and for (effective) Quantum General Relativity via \(\mathcal{T}_\text{QGR} \coloneq \set{\mathbbit{L}, \mathbbit{I}, \mathbbit{T}}_{\mathbbit{\mettransstr}}\) with \emph{physical projection operator} \(\mathbbit{P}\).
\end{defn}

\enter

\begin{defn}[Optimal gauge fixing] \label{defn:optimal-gauge-fixing}
	Let \(\Q\) be a quantum gauge theory with Lagrange density
	\begin{equation}
		\LQ = \mathcal{L}_\text{Classical} + \mathcal{L}_\text{GF} + \mathcal{L}_\text{Ghost} \, .
	\end{equation}
	Then, the tensor \(\boldsymbol{T}\) is proportional to the Feynman rule of the quadratic term in \(\mathcal{L}_\text{Classical}\). Furthermore, the tensor \(\boldsymbol{L}\) is proportional to the Feynman rule of the quadratic term in \(\mathcal{L}_\text{GF}\). Finally, we call a gauge fixing functional \emph{optimal} if these tensors satisfy
	\begin{equation}
		\boldsymbol{T} + \boldsymbol{L} = \boldsymbol{\mettransstr} \, ,
	\end{equation}
	where \(\boldsymbol{\mettransstr}\) denotes the corresponding metric tensor. Equivalently, the quadratic Feynman rule of the gauge boson is proportional to \(\bigl ( \boldsymbol{T} + \boldsymbol{L} / \varrho \bigr )\) and its propagator is proportional to \(\bigl ( \boldsymbol{T} + \varrho \boldsymbol{L} \bigr )\), where \(\varrho\) denotes the gauge fixing parameter.
\end{defn}

\enter

\begin{rem}
	The notion of an \emph{optimal gauge fixing} in \defnref{defn:optimal-gauge-fixing} corresponds to the fact that it only adds an evolutionary equation to the gauge degrees of freedom and leaves the evolutionary equation for the physical degrees of freedom unchanged. Specifically, this means that \(\mathcal{L}_\text{Classical} + \mathcal{L}_\text{GF} \big \vert_{\varrho = 1}\) corresponds to the metric tensor for the corresponding Lorentz tensor structure. A prominent example of a non-optimal gauge fixing is the Curci--Ferrari gauge \cite{Curci_Ferrari}, which is in particular non-linear and thus leads to four-valent ghost vertices.
\end{rem}

\subsection{Quantum Yang--Mills theory with matter} \label{ssec:transversality_qym}

We recall known and immediate identities for the transversal structure of Quantum Yang--Mills theory with a Lorenz gauge fixing. To this end, we refer the reader again to \sectionref{sec:explicit_feynman_rules} for details on particle flow, momentum flow and index placement of the Feynman rules, in order to avoid a notational overload here.

\enter

\begin{defn}[Transversal structure of QYM] \label{defn:qym-transversal-structure}
	Consider Quantum Yang--Mills theory with a Lorenz gauge fixing. Then we set its transversal structure \(\mathcal{T}_\text{QYM} \coloneq \setbig{L, I, T}_\mettransstr\) as follows:\footnote{We remark that the color indices are implicitly included in the tensors \(L\), \(I\) and \(T\) by considering their tensor product with the identity matrix \(\delta\). We suppress this to simplify the notation.}
	\begin{subequations} \label{eqn:defn_projection_tensors_qym}
	\begin{align}
		L^\nu_\mu & \coloneq \frac{1}{p^2} p^\nu p_\mu \, , \\
		I^\nu_\mu & \coloneq \delta^\nu_\mu
		\intertext{and}
		T^\nu_\mu & \coloneq I^\nu_\mu - L^\nu_\mu \, , \label{eqn:defn-t-qym}
	\end{align}
	\end{subequations}
	where we have set \(p^2 \coloneq \eta_{\mu \nu} p^\mu p^\nu\). Lorentz indices on \(L\), \(I\) and \(T\) are raised and lowered with the metric \(\mettransstr\), defined via\footnote{We note the additional factors of \(1 / p^2\) and \(p^2\) in the definitions of \(\mettransstr_{\mu \nu}\) and \(\mettransstr^{\mu \nu}\), respectively, which will be convenient in \thmref{thm:feynman-rule-gluon-propagator-lt-decomposition}.}
	\begin{subequations}
	\begin{align}
		\mettransstr_{\mu \nu} & \coloneq \frac{1}{p^2} \eta_{\mu \nu}
		\intertext{and its inverse}
		\mettransstr^{\mu \nu} & \coloneq p^2 \eta^{\mu \nu} \, .
	\end{align}
	\end{subequations}
	Finally, we define the following three (co)vectors
	\begin{subequations} \label{eqns:gauge_transformation_and_longitudinal_projection_tensors_qym}
	\begin{align}
		\gtym_\mu & \coloneq \frac{1}{p^2} p_\mu \, ,
		\intertext{its prefactor reduced variant \(\gtym^\prime_\mu \coloneq p^2 \gtym_\mu\) for well-defined on-shell contraction identities, and}
		\gfym^\nu & \coloneq p^\nu \, .
	\end{align}
	\end{subequations}
\end{defn}

\enter

\begin{rem}
	The tensor \(\gtym\) corresponds to a linearized gauge transformation (sometimes also called \emph{gaugeon projection}) and the tensor \(\gfym\) describes the gauge fixing projection for the Lorenz gauge. Furthermore, their degree in \(p^2\) is chosen such that the contraction with \(\gtym\) corresponds to the contraction with half of a longitudinal gauge boson propagator.
\end{rem}

\enter

\begin{lem} \label{lem:g_and_f_inverse_decomposition_gl_qym}
	The following identities hold, i.e.\ \(\gtym\) and \(\gfym\) are inverse to each other and \(L\) decomposes into the product of \(\gtym\) and \(\gfym\):
	\begin{subequations}
	\begin{align}
		\gtym_\mu \gfym^\mu & = 1 \label{eqn:g_and_l_inverse_qym} \\
		\gfym^\nu \gtym_\mu & = L^\nu_\mu \label{eqn:decomposition_longitudinal_projection_tensor_qym}
	\end{align}
	\end{subequations}
\end{lem}

\begin{proof}
	This follows immediately from basic tensor calculations.
\end{proof}

\enter

\begin{lem} \label{lem:identities_tensors_qym}
	The following identities hold, i.e.\ \(\gtym\) and \(\gfym\) are related via inversion with respect to \(\mettransstr\):
	\begin{subequations}
	\begin{align}
		\mettransstr_{\mu \nu} \gfym^\nu & = \gtym_\mu \label{eqn:g_and_l_relation_metric_qym} \\
		\mettransstr^{\mu \nu} \gtym_\mu & = \gfym^\nu \label{eqn:g_and_l_relation_inverse-metric_qym}
	\end{align}
	\end{subequations}
\end{lem}

\begin{proof}
	This follows immediately from basic tensor calculations.
\end{proof}

\enter

\begin{col} \label{col:l-tensor-gg-ff-qym}
	The following identities hold, i.e.\ \(L\) with raised and lowered indices decomposes into products of two \(\gtym\) or \(\gfym\) (co)vectors, respectively:
	\begin{subequations}
	\begin{align}
		L_{\mu \nu} & = \gtym_\mu \gtym_\nu \label{eqn:l_gg_qym} \\
		L^{\mu \nu} & = \gfym^\mu \gfym^\nu \label{eqn:l_ll_inverse_qym}
	\end{align}
	\end{subequations}
\end{col}

\begin{proof}
	This follows immediately from \lemsaref{lem:g_and_f_inverse_decomposition_gl_qym}{lem:identities_tensors_qym}.
\end{proof}

\enter

\begin{prop} \label{prop:gf-eigenvectors-lit-qym}
	The two (co)vectors \(\gtym\) and \(\gfym\) are eigen(co)vectors of the tensors \(L\), \(I\) and \(T\) with respective eigenvalues \(1\) and \(0\), i.e.\ we have:
	\begin{subequations}
	\begin{align}
		L^\nu_\mu \gtym_\nu & = \gtym_\mu \label{eqn:g_eigentensor_qym} \\
		L^\nu_\mu \gfym^\mu & = \gfym^\nu \label{eqn:f_eigentensor_qym} \\
		I^\nu_\mu \gtym_\nu & = \gtym_\mu \label{eqn:g_identity_qym} \\
		I^\nu_\mu \gfym^\mu & = \gfym^\nu \label{eqn:f_identity_qym} \\
		T^\nu_\mu \gtym_\nu & = 0 \label{eqn:g_orthogonal_qym} \\
		T^\nu_\mu \gfym^\mu & = 0 \label{eqn:f_orthogonal_qym}
	\end{align}
	\end{subequations}
\end{prop}

\begin{proof}
	This follows immediately from \lemref{lem:g_and_f_inverse_decomposition_gl_qym} and basic tensor calculations.
\end{proof}

\enter

\begin{thm} \label{thm:qym-transversal-structure}
	The following identities hold, i.e.\ the tensors \(L\), \(I\) and \(T\) are projectors:
	\begin{subequations}
	\begin{align}
		L_\mu^\tau L_\tau^\nu & = L_\mu^\nu \\
		I_\mu^\tau I_\tau^\nu & = I_\mu^\nu \\
		T_\mu^\tau T_\tau^\nu & = T_\mu^\nu \\
		L_\mu^\tau I_\tau^\nu & = L_\mu^\nu \\
		T_\mu^\tau I_\tau^\nu & = T_\mu^\nu \\
		L_\mu^\tau T_\tau^\nu & = 0
	\end{align}
	\end{subequations}
	Additionally, the tensors \(L\) and \(T\) are orthogonal and the tensor \(I\) is the identity with respect to the metric \(\mettransstr\) and its inverse \(\mettransstr^{-1}\):
	\begin{equation}
		\mettransstr_{\mu \tau} \mettransstr^{\tau \nu} = I^\nu_\mu
	\end{equation}
\end{thm}

\begin{proof}
	The idempotency follows immediately from \lemref{lem:g_and_f_inverse_decomposition_gl_qym} together with \propref{prop:gf-eigenvectors-lit-qym}. Further, the orthogonality and identity property then follow in addition with \colref{col:l-tensor-gg-ff-qym}.
\end{proof}

\enter

\begin{thm} \label{thm:feynman-rule-gluon-propagator-lt-decomposition}
	The Feynman rule of the gauge boson propagator can be written as follows:
	\begin{equation} \label{eqn:decomposition_gluon_propagator}
		\begin{split}
			\Phi \left ( \cgreen{p-gluon} \right ) & = - \frac{\imaginary p^2}{p^2 + \imaginary \epsilon} \delta^{a b} \left ( \mettransstr_{\mu \nu} - \left ( 1 - \xi \right ) L_{\mu \nu} \right ) \\
			& = - \frac{\imaginary p^2}{p^2 + \imaginary \epsilon} \delta^{a b} \left ( T_{\mu \nu} + \xi L_{\mu \nu} \right )
		\end{split}
	\end{equation}
	Furthermore, the Feynman rules of the gauge boson propagator and the gauge ghost propagator are related as follows, where the vector arrows indicate which legs are contracted with the indices of the (co)vectors \(\gfym\) and \(\gtym\):
	\begin{subequations}\label{eqn:relations_gluon_gluon-ghost_propagators}
	\begin{align}
		& \Phi \bigl ( {\scriptstyle \vec{\gfym}} \cgreen{p-gluon} \bigr ) = p^2 \Phi \bigl ( \cgreen{p-gluonghost} {\scriptstyle \vec{\gtym}} \bigr ) \, , \\[0.5\baselineskip]
		& \Phi \bigl ( {\scriptstyle \vec{\gfym}} \cgreen{p-gluon} {\scriptstyle \cevss{\gfym}} \bigr ) = p^2 \Phi \bigl ( \cgreen{p-gluonghost} \bigr )
		\intertext{and}
		& \Phi \bigl ( {\scriptstyle L} \cgreen{p-gluon} {\scriptstyle L} \bigr ) = p^2 \Phi \bigl ( {\scriptstyle \cevss{\gtym}} \cgreen{p-gluonghost} {\scriptstyle \vec{\gtym}} \bigr ) \, .
	\end{align}
	\end{subequations}
	In particular, the Lorenz gauge fixing is the optimal gauge fixing condition for Quantum Yang--Mills theory.
\end{thm}

\begin{proof}
	\eqnref{eqn:decomposition_gluon_propagator} follows from the previous results in this subsection together with the Feynman rule
	\begin{align}
		\Phi \left ( \cgreen{p-gluon} \right ) & = - \frac{\imaginary}{p^2 + \imaginary \epsilon} \delta^{a b} \left ( \eta_{\mu \nu} - \frac{\left ( 1 - \xi \right )}{p^2} p_\mu p_\nu \right ) \, .
		\intertext{From this, \eqnsref{eqn:relations_gluon_gluon-ghost_propagators} follows from the previous results in this subsection together with the Feynman rule}
		\Phi \left ( \cgreen{p-gluonghost} \right ) & = - \frac{\imaginary \xi}{p^2 + \imaginary \epsilon} \delta^{a b} \, .
	\end{align}
	The final claim follows then directly from \eqnref{eqn:decomposition_gluon_propagator}.
\end{proof}

\enter

\begin{thm} \label{thm:three-valent-contraction-identities-qym}
	The Feynman rule of the three-valent gauge boson vertex satisfies the following contraction identities:
	{\allowdisplaybreaks
	\begin{subequations}
	\begin{align}
		& \Phi \left ( {\scriptstyle \gtym} \tcgreen{v-gluontriple} _{{\scriptstyle \gtym}}^{{\scriptstyle \gtym}} \rule{0pt}{5ex} \right ) = \Phi \left ( {\scriptstyle L} \tcgreen{v-gluontriple} _{{\scriptstyle L}}^{{\scriptstyle L}} \rule{0pt}{5ex} \right ) = 0 \label{eqn:contraction_v-triple-gluon} \, , \\
		& \Phi \left ( {\scriptstyle \gtymred} \tcgreen{v-gluontriple} _{{\scriptstyle P}}^{{\scriptstyle P}} \rule{0pt}{5ex} \right ) = 0 \label{eqn:contraction_v-triple-gluon-os-tt}
		\intertext{and thus}
		& \Phi \left ( {\scriptstyle \gtymred} \tcgreen{v-gluontriple} _{{\scriptstyle I}}^{{\scriptstyle I}} \rule{0pt}{5ex} \right ) \simeq_\textup{OS} \Phi \left ( {\scriptstyle \gtymred} \tcgreen{v-gluontriple} _{{\scriptstyle L}}^{{\scriptstyle T}} \rule{0pt}{5ex} \right ) + \Phi \left ( {\scriptstyle \gtymred} \tcgreen{v-gluontriple} _{{\scriptstyle T}}^{{\scriptstyle L}} \rule{0pt}{5ex} \right ) \, , \label{eqn:contraction_v-triple-gluon-os-ii}
	\end{align}
	\end{subequations}
	}%
	where \(P\) denotes the projector onto the physical degrees of freedom, i.e.\ first the on-shell reduction by setting \(p^2 = 0\), and then the transversal projection via \(T\). In addition, \(\simeq_\textup{OS}\) indicates equality on-shell.
\end{thm}

\begin{proof}
	Starting with the first identity, we recall the decomposition \(L_\mu^\nu = \gfym^\nu \gtym_\mu\) of \eqnref{eqn:decomposition_longitudinal_projection_tensor_qym} and therefore only calculate the contraction with the \(\gtym\) covectors:
	\begin{equation}
	\begin{split}
		\Phi \left ( {\scriptstyle \gtym} \tcgreen{v-gluontriple} _{{\scriptstyle \gtym}}^{{\scriptstyle \gtym}} \rule{0pt}{5ex} \right ) & = - \mathrm{g} f_{a_1 a_2 a_3} \frac{1}{p_1^2 \, p_2^2 \, p_3^2} p^1_{\rho_1} p^2_{\rho_2} p^3_{\rho_3} \\ & \phantom{= -} \times \biggl ( \eta^{\rho_1 \rho_2} \left ( p_1 - p_2 \right )^{\rho_3} + \eta^{\rho_2 \rho_3} \left ( p_2 - p_3 \right )^{\rho_1} + \eta^{\rho_3 \rho_1} \left ( p_3 - p_1 \right )^{\rho_2} \biggr ) \vphantom{\frac{1}{p_1^2 \, p_2^2 \, p_3^2}} \\
	& = 0 \vphantom{\frac{1}{p_1^2 \, p_2^2 \, p_3^2}}
	\end{split}
	\end{equation}
	For the two remaining identities, we remind the reader of the reduced gauge transformation contraction \(\gtymred_\mu \coloneq p^2 \gtym_\mu\), recall the decomposition \(I_\mu^\nu = T_\mu^\nu + L_\mu^\nu\) of \eqnref{eqn:defn-t-qym} and calculate
	\begin{equation}
	\begin{split}
		\Phi \left ( {\scriptstyle \gtymred} \tcgreen{v-gluontriple} _{{\scriptstyle I}}^{{\scriptstyle I}} \rule{0pt}{5ex} \right ) & = - \mathrm{g} f_{a_1 a_2 a_3} p^1_{\rho_1} \\ & \phantom{= -} \times \biggl ( \eta^{\rho_1 \rho_2} \left ( p_1 - p_2 \right )^{\rho_3} + \eta^{\rho_2 \rho_3} \left ( p_2 - p_3 \right )^{\rho_1} + \eta^{\rho_3 \rho_1} \left ( p_3 - p_1 \right )^{\rho_2} \biggr ) \vphantom{\frac{1}{p_1^2}} \\
	& \simeq_\text{MC} \mathrm{g} f_{a_1 a_2 a_3} \left ( I^{\rho_2 \rho_3} \left ( p_2^\sigma \right ) - I^{\rho_2 \rho_3} \left ( p_3^\sigma \right ) - L^{\rho_2 \rho_3} \left ( p_2^\sigma \right ) + L^{\rho_2 \rho_3} \left ( p_3^\sigma \right ) \right ) \vphantom{\Big \vert} \\
	 & \simeq_\text{OS} - \mathrm{g} f_{a_1 a_2 a_3} \left ( L^{\rho_2 \rho_3} \left ( p_2^\sigma \right ) - L^{\rho_2 \rho_3} \left ( p_3^\sigma \right ) \right ) \vphantom{\Big \vert}
	\end{split}
	\end{equation}
	by noting \(I^{\mu \nu} = p^2 \eta^{\mu \nu}\) and recalling the identity \(L^{\mu \nu} \, T_\mu^\rho = 0\). In addition, \(\simeq_\text{MC}\) indicates equality modulo momentum conservation, i.e.\ setting \(p^\sigma \coloneq - q_1^\sigma - q_2^\sigma\), and \(\simeq_\text{OS}\) indicates equality on-shell, i.e.\ setting \(q_1^2 = q_2^2 = 0\).
\end{proof}

\enter

\begin{rem} \label{rem:contraction-single-gluon}
	\eqnsaref{eqn:contraction_v-triple-gluon-os-tt}{eqn:contraction_v-triple-gluon-os-ii} imply that the longitudinal projection of a single gauge boson results in both transversal on-shell cancellations and propagating longitudinal gauge boson modes. We will find out that this is similar in (effective) Quantum General Relativity, cf.\ \remref{rem:contraction-single-graviton}.
\end{rem}

\enter

\begin{thm} \label{thm:three-valent-contraction-identities-qym-with-matter}
	The Feynman rules of the three-valent interactions of gauge bosons with scalars, spinors and gauge ghosts satisfy the following on-shell contraction identities:
	{\allowdisplaybreaks
	\begin{align}
		& \Phi \left ( {\scriptstyle \gtymred} \tcgreen{v-gluonscalartriple}^{{\scriptstyle \OS}}_{{\scriptstyle \OS}} \right ) = 0 \label{eqn:contraction_v-gluonscalartriple} \\
		& \Phi \left ( {\scriptstyle \gtymred} \tcgreen{v-gluonspinortriple}^{{\scriptstyle \OS}}_{{\scriptstyle \OS}} \right ) = 0 \label{eqn:contraction_v-gluonspinortriple} \\
		& \Phi \left ( {\scriptstyle \gtymred} \tcgreen{v-gluonghosttriple}^{{\scriptstyle \OS}}_{{\scriptstyle \OS}} \right ) = 0 \label{eqn:contraction_v-gluonghosttriple}
	\end{align}
	}%
	and with the corresponding off-shell variants leading to the usual propagator cancellations, as can be seen in the calculations of the proof.
\end{thm}

\begin{proof}
	Again, we consider all momenta incoming and denote the gauge boson momentum by \(p^\sigma\) and the matter momenta by \(q_1^\sigma\) and \(q_2^\sigma\). Also, we denote the gauge boson Lorentz and color indices by \(\rho\) and \(a\), respectively. Moreover, \(\mathfrak{H}\) and \(\mathfrak{S}\) denote the infinitesimal gauge group actions on the Higgs bundle and spinor bundle, respectively. In \eqnsaref{eqn:proof_contraction_v-gluonspinortriple}{eqn:proof_contraction_v-gluonghosttriple} the number 1 corresponds to the particle and the number 2 to the anti-particle. In particular, in \eqnref{eqn:proof_contraction_v-gluonspinortriple} this implies that the equations of motion differ in a relative sign. In addition, in \eqnref{eqn:proof_contraction_v-gluonghosttriple} we denote the gauge ghost color indices by \(c_1\) and \(c_2\), respectively. In \eqnref{eqn:proof_contraction_v-gluonghosttriple} we use the symmetric (hermitian) gauge ghost Lagrange density of \eqnref{eqn:sym-gf-ghost-ym}. With that, we present the actual calculations (the appearing masses \(m\) correspond to the respective matter particles):
	{\allowdisplaybreaks
	\begin{align}
	\begin{split} \label{eqn:proof_contraction_v-gluonscalartriple}
		\Phi \left ( {\scriptstyle \gtymred} \tcgreen{v-gluonscalartriple} \right ) & = - \mathrm{g} \, p_\rho \biggl ( \left ( q_1 - q_2 \right )^\rho \left ( \mathfrak{H}_a \right )_{i j} \biggr ) \\
		& \simeq_\text{MC} \mathrm{g} \left ( q_1^2 - q_2^2 \right ) \left ( \mathfrak{H}_a \right )_{i j} \\
		& = \mathrm{g} \left ( \left ( q_1^2 - m^2 \right ) - \left ( q_2^2 - m^2 \right ) \right ) \left ( \mathfrak{H}_a \right )_{i j} \\
		& \simeq_\text{OS} 0 \vphantom{\bigg \vert}
		\end{split} \\
		\begin{split} \label{eqn:proof_contraction_v-gluonspinortriple}
		\Phi \left ( {\scriptstyle \gtymred} \tcgreen{v-gluonspinortriple} \right ) & = - \imaginary \mathrm{g} \, p_\rho \left ( \gamma^\rho \left ( \mathfrak{S}_a \right )_{k l} \right ) \\
		& \simeq_\text{MC} \imaginary \mathrm{g} \left ( \left ( \slashed{q}_1 - m \right ) + \left ( \slashed{q}_2 + m \right ) \right ) \left ( \mathfrak{S}_a \right )_{k l} \\
		& \simeq_\text{OS} 0 \vphantom{\bigg \vert}
		\end{split} \\
		\begin{split} \label{eqn:proof_contraction_v-gluonghosttriple}
		\Phi \left ( {\scriptstyle \gtymred} \tcgreen{v-gluonghosttriple} \right ) & = \frac{\mathrm{g}}{2} \, p_\rho \left ( f_{a b c} \left ( q_1 - q_2 \right )^\rho \right ) \\
		& \simeq_\text{MC} - \frac{\mathrm{g}}{2} f_{a b c} \left ( q_1^2 - q_2^2 \right ) \\
		& \simeq_\text{OS} 0 \vphantom{\bigg \vert} \, ,
		\end{split}
	\end{align}
	}%
	where \(\simeq_\text{MC}\) indicates equality modulo momentum conservation, i.e.\ setting \(p^\sigma \coloneq - q_1^\sigma - q_2^\sigma\), and \(\simeq_\text{OS}\) indicates equality on-shell, i.e.\ setting \(q_1^2 = q_2^2 = 0\).
\end{proof}

\subsection{(Effective) Quantum General Relativity with matter} \label{ssec:transversality_qgr}

We introduce novel and involved identities for the transversal structure of (effective) Quantum General Relativity with a de Donder gauge fixing. To this end, we refer the reader again to \sectionref{sec:explicit_feynman_rules} for details on particle flow, momentum flow and index placement of the Feynman rules, in order to avoid a notational overload here.

\enter

\begin{defn}[Transversal structure of QGR] \label{defn:qgr-transversal-structure}
	Consider (effective) Quantum General Relativity with a de Donder gauge fixing. Then we set its transversal structure \(\mathcal{T}_\text{QGR} \coloneq \setbig{\mathbbit{L}, \mathbbit{I}, \mathbbit{T}}_{\mathbbit{\mettransstr}}\) as follows:
	\begin{subequations} \label{eqn:defn_projection_tensors_qgr}
	\begin{align}
		\mathbbit{L}^{\rho \sigma}_{\mu \nu} & \coloneq \frac{1}{2 p^2} \left ( \delta^\rho_\mu p^\sigma p_\nu + \delta^\sigma_\mu p^\rho p_\nu + \delta^\rho_\nu p^\sigma p_\mu + \delta^\sigma_\nu p^\rho p_\mu - 2 \eta^{\rho \sigma} p_\mu p_\nu \right ) \, , \\
		\mathbbit{I} \mspace{2mu} ^{\rho \sigma}_{\mu \nu} & \coloneq \frac{1}{2} \left ( \delta^\rho_\mu \delta^\sigma_\nu + \delta^\sigma_\mu \delta^\rho_\nu \right )
		\intertext{and}
		\mathbbit{T} \mspace{2mu} ^{\rho \sigma}_{\mu \nu} & \coloneq \mathbbit{I} \mspace{2mu} ^{\rho \sigma}_{\mu \nu} - \mathbbit{L}^{\rho \sigma}_{\mu \nu} \, , \label{eqn:defn-t-qgr}
	\end{align}
	\end{subequations}
	where we have set \(p^2 \coloneq \eta_{\mu \nu} p^\mu p^\nu\). Lorentz indices on \(\mathbbit{L}\), \(\mathbbit{I}\) and \(\mathbbit{T}\) are raised and lowered with the metric \(\mathbbit{\mettransstr}\), defined via\footnote{We note the additional factors of \(1 / p^2\) and \(p^2\) in the definitions of \(\mathbbit{\mettransstr}_{\mu \nu \rho \sigma}\) and \(\mathbbit{\mettransstr}^{\mu \nu \rho \sigma}\), respectively, which will be convenient in \thmref{thm:feynman-rule-graviton-propagator-lt-decomposition}. Furthermore, we emphasize that the asymmetric definition concerning the factor \(\textfrac{1}{4}\) is motivated by \eqnsref{eqns:gauge_transformation_and_longitudinal_projection_tensors_qgr}. Moreover, we remark that for a general spacetime dimension \(d\), \eqnref{eqn:grav-trans-metric} actually reads \begin{equation*} \mathbbit{\mettransstr}_{\mu \nu \rho \sigma} \coloneq \frac{1}{p^2} \left ( \eta_{\mu \rho} \eta_{\nu \sigma} + \eta_{\mu \sigma} \eta_{\nu \rho} - \frac{2}{d - 2} \eta_{\mu \nu} \eta_{\rho \sigma} \right )\end{equation*} with \eqnref{eqn:grav-trans-invmetric} unchanged. This also affects the graviton propagator, cf.\ \thmref{thm:feynman-rule-graviton-propagator-lt-decomposition}. However, for the present article we stick to \(d = 4\) and discuss the situation for general spacetime dimension \(d\) --- e.g.\ relevant for dimensional regularization --- more detailed in future work.}
	\begin{subequations}
	\begin{align}
		\mathbbit{\mettransstr}_{\mu \nu \rho \sigma} & \coloneq \frac{1}{p^2} \left ( \eta_{\mu \rho} \eta_{\nu \sigma} + \eta_{\mu \sigma} \eta_{\nu \rho} - \eta_{\mu \nu} \eta_{\rho \sigma} \right ) \label{eqn:grav-trans-metric}
		\intertext{and its inverse}
		\mathbbit{\mettransstr}^{\mu \nu \rho \sigma} & \coloneq \frac{p^2}{4} \left ( \eta^{\mu \rho} \eta^{\nu \sigma} + \eta^{\mu \sigma} \eta^{\nu \rho} - \eta^{\mu \nu} \eta^{\rho \sigma} \right ) \, . \label{eqn:grav-trans-invmetric}
	\end{align}
	\end{subequations}
	Finally, we define the following three tensors
	\begin{subequations} \label{eqns:gauge_transformation_and_longitudinal_projection_tensors_qgr}
	\begin{align}
		\gtgr_{\mu \nu}^\kappa & \coloneq \frac{1}{p^2} \bigl ( p_\mu \delta_\nu^\kappa + p_\nu \delta_\mu^\kappa \bigr ) \, ,
		\intertext{its prefactor reduced variant \(\gtgr_{\mu \nu}^{\prime \kappa} \coloneq p^2 \gtgr_{\mu \nu}^\kappa\) for well-defined on-shell contraction identities, and}
		\gfgr^{\rho \sigma}_\lambda & \coloneq \frac{1}{2} \bigl ( p^\rho \delta^\sigma_\lambda + p^\sigma \delta^\rho_\lambda - p_\lambda \eta^{\rho \sigma} \bigr ) \, .
	\end{align}
	\end{subequations}
\end{defn}

\enter

\begin{rem}
	The tensor \(\gtgr\) corresponds to a linearized gauge transformation (sometimes also called \emph{gaugeon projection}) and the tensor \(\gfgr\) describes the gauge fixing projection for the de Donder gauge. Furthermore, their degree in \(p^2\) is chosen such that the contraction with \(\gtgr\) corresponds to the contraction with half of a longitudinal graviton propagator. Additionally, we emphasize that --- unlike in Quantum Yang--Mills theory, cf.\ \defnref{defn:qym-transversal-structure} --- the tensors \(\mathbbit{L}\) and \(\mathbbit{T}\) are \emph{not} symmetric with respect to their lower and upper index pairs: Thus, they are \emph{oblique} projection tensors with respect to the \emph{naive} metric \(\widetilde{\mathbbit{\mettransstr}}_{\mu \nu \rho \sigma} \coloneq \textfrac{\left ( \eta_{\mu \rho} \eta_{\nu \sigma} + \eta_{\mu \sigma} \eta_{\nu \rho} \right )}{p^2}\). However, as we will see in \colref{col:l-tensor-gg-ff-qgr}, they are orthogonal with respect to the \emph{transversal structure} metric \(\mathbbit{\mettransstr}_{\mu \nu \rho \sigma}\) from \eqnref{eqn:grav-trans-metric}. The reason is that the Einstein--Hilbert Lagrange density has tensor density weight \(w = 1\) and thus the graviton field transforms differently from the remaining matrix element, which gives rise to the vertex Feynman rules. Notably, \(\gtgr\) corresponds to a linearized gauge transformation of a \((0,2)\)-tensor of weight \(w = 0\), while \(\gfgr\) corresponds to a linearized gauge transformation of a \((2,0)\)-tensor of weight \(w = 1\). Further, we note that the metric \(\mathbbit{E}\) is proportional to the \emph{DeWitt metric}, with proportionality constant \(\textfrac{p^2}{2}\), \cite{DeWitt_I,DeWitt_II,DeWitt_III,DeWitt_IV}, cf.\ \cite{Giulini_Kiefer}. Finally, we emphasize that physical gravitons are additionally \emph{traceless}: We omit this fact in the present article, as we are interested in the \emph{diagrammatic cancellation identities}, which work with \emph{on-shell projections} and \emph{longitudinal modes}.
\end{rem}

\enter

\begin{lem} \label{lem:g_and_f_inverse_decomposition_gl_qgr}
	The following identities hold, i.e.\ \(\gtgr\) and \(\gfgr\) are inverse to each other and \(\mathbbit{L}\) decomposes into the product of \(\gtgr\) and \(\gfgr\):
	\begin{subequations}
	\begin{align}
		\gtgr_{\mu \nu}^\kappa \gfgr^{\mu \nu}_\lambda & = \delta^\kappa_\lambda \label{eqn:g_and_l_inverse_qgr} \\
		\gfgr^{\rho \sigma}_\tau \gtgr_{\mu \nu}^\tau & = \mathbbit{L}^{\rho \sigma}_{\mu \nu} \label{eqn:decomposition_longitudinal_projection_tensor_qgr}
	\end{align}
	\end{subequations}
\end{lem}

\begin{proof}
	This follows immediately from basic tensor calculations.
\end{proof}

\enter

\begin{lem} \label{lem:identities_tensors_qgr}
	The following identities hold, i.e.\ \(\gtgr\) and \(\gfgr\) are related via inversion with respect to \(\mathbbit{\mettransstr} \otimes \eta\):
	\begin{subequations}
	\begin{align}
		\mathbbit{\mettransstr}_{\mu \nu \rho \sigma} \eta^{\kappa \lambda} \gfgr^{\rho \sigma}_\lambda & = \gtgr_{\mu \nu}^\kappa \label{eqn:g_and_l_relation_metric_qgr} \\
		\mathbbit{\mettransstr}^{\mu \nu \rho \sigma} \eta_{\kappa \lambda} \gtgr_{\mu \nu}^\kappa & = \gfgr^{\rho \sigma}_\lambda \label{eqn:g_and_l_relation_inverse-metric_qgr}
	\end{align}
	\end{subequations}
\end{lem}

\begin{proof}
	This follows immediately from basic tensor calculations.
\end{proof}

\enter

\begin{col} \label{col:l-tensor-gg-ff-qgr}
	The following identities hold, i.e.\ \(\mathbbit{L}\) with raised and lowered indices decomposes into products of two \(\gtgr\) or \(\gfgr\) tensors, respectively. In particular, the tensors \(\mathbbit{L}\) and \(\mathbbit{T}\) are self-adjoint, i.e.\ symmetric under the exchange \(\set{\mu \nu} \leftrightarrow \set{\rho \sigma}\), with respect to the metric \(\mathbbit{\mettransstr}\):
	\begin{subequations}
	\begin{align}
		\mathbbit{L}_{\mu \nu \rho \sigma} & = \eta_{\kappa \lambda} \gtgr_{\mu \nu}^\kappa \gtgr_{\rho \sigma}^\lambda \label{eqn:l_gg_qgr} \\
		\mathbbit{L}^{\mu \nu \rho \sigma} & = \eta^{\kappa \lambda} \gfgr^{\mu \nu}_\kappa \gfgr^{\rho \sigma}_\lambda \label{eqn:l_ll_inverse_qgr}
	\end{align}
	\end{subequations}
\end{col}

\begin{proof}
	The statement about \(\mathbbit{L}\) follows immediately from \lemsaref{lem:g_and_f_inverse_decomposition_gl_qgr}{lem:identities_tensors_qgr}. Additionally, the statement for \(\mathbbit{T}\) then follows directly due to the respective symmetries of \(\mathbbit{I}\) and \(\mathbbit{\mettransstr}\).
\end{proof}

\enter

\begin{prop} \label{prop:gf-eigentensors-lit-qgr}
	The two tensors \(\gtgr\) and \(\gfgr\) are eigentensors of the tensors \(\mathbbit{L}\), \(\mathbbit{I}\) and \(\mathbbit{T}\) with respective eigenvalues \(1\) and \(0\), i.e.\ we have:
	\begin{subequations}
	\begin{align}
		\mathbbit{L}^{\rho \sigma}_{\mu \nu} \gtgr_{\rho \sigma}^\kappa & = \gtgr_{\mu \nu}^\kappa \label{eqn:g_eigentensor_qgr} \\
		\mathbbit{L}^{\rho \sigma}_{\mu \nu} \gfgr^{\mu \nu}_\lambda & = \gfgr^{\rho \sigma}_\lambda \label{eqn:f_eigentensor_qgr} \\
		\mathbbit{I}^{\rho \sigma}_{\mu \nu} \gtgr_{\rho \sigma}^\kappa & = \gtgr_{\mu \nu}^\kappa \label{eqn:g_identity_qgr} \\
		\mathbbit{I}^{\rho \sigma}_{\mu \nu} \gfgr^{\mu \nu}_\lambda & = \gfgr^{\rho \sigma}_\lambda \label{eqn:f_identity_qgr} \\
		\mathbbit{T}^{\rho \sigma}_{\mu \nu} \gtgr_{\rho \sigma}^\kappa & = 0 \label{eqn:g_orthogonal_qgr} \\
		\mathbbit{T}^{\rho \sigma}_{\mu \nu} \gfgr^{\mu \nu}_\lambda & = 0 \label{eqn:f_orthogonal_qgr}
	\end{align}
	\end{subequations}
\end{prop}

\begin{proof}
	This follows immediately from \lemref{lem:g_and_f_inverse_decomposition_gl_qgr} and basic tensor calculations.
\end{proof}

\enter

\begin{thm} \label{thm:qgr-transversal-structure}
	The following identities hold, i.e.\ the tensors \(\mathbbit{L}\), \(\mathbbit{I}\) and \(\mathbbit{T}\) are projectors:
	\begin{subequations}
	\begin{align}
		\mathbbit{L}^{\kappa \lambda}_{\mu \nu} \, \mathbbit{L}^{\rho \sigma}_{\kappa \lambda} & = \mathbbit{L}^{\rho \sigma}_{\mu \nu} \\
		\mathbbit{I}^{\kappa \lambda}_{\mu \nu} \, \mathbbit{I}^{\rho \sigma}_{\kappa \lambda} & = \mathbbit{I}^{\rho \sigma}_{\mu \nu} \\
		\mathbbit{T}^{\kappa \lambda}_{\mu \nu} \, \mathbbit{T}^{\rho \sigma}_{\kappa \lambda} & = \mathbbit{T}^{\rho \sigma}_{\mu \nu} \\
		\mathbbit{L}^{\kappa \lambda}_{\mu \nu} \, \mathbbit{I} \mspace{1mu} ^{\rho \sigma}_{\kappa \lambda} & = \mathbbit{L}^{\rho \sigma}_{\mu \nu} \\
		\mathbbit{T}^{\kappa \lambda}_{\mu \nu} \, \mathbbit{I} \mspace{1mu} ^{\rho \sigma}_{\kappa \lambda} & = \mathbbit{T}^{\rho \sigma}_{\mu \nu} \\
		\mathbbit{L}^{\kappa \lambda}_{\mu \nu} \, \mathbbit{T}^{\rho \sigma}_{\kappa \lambda} &= 0
	\end{align}
	\end{subequations}
	Additionally, the tensors \(\mathbbit{L}\) and \(\mathbbit{T}\) are orthogonal and the tensor \(\mathbbit{I}\) is the identity with respect to the metric \(\mathbbit{\mettransstr}\) and its inverse \(\mathbbit{\mettransstr}^{-1}\):
	\begin{equation}
		\mathbbit{\mettransstr}_{\mu \nu \kappa \lambda} \mathbbit{\mettransstr}^{\kappa \lambda \rho \sigma} = \mathbbit{I} \mspace{2mu} ^{\rho \sigma}_{\mu \nu}
	\end{equation}
\end{thm}

\begin{proof}
	The idempotency follows immediately from \lemref{lem:g_and_f_inverse_decomposition_gl_qgr} together with \propref{prop:gf-eigentensors-lit-qgr}. Further, the orthogonality and identity property then follow in addition with \colref{col:l-tensor-gg-ff-qgr}.
\end{proof}

\enter

\begin{thm} \label{thm:feynman-rule-graviton-propagator-lt-decomposition}
	The Feynman rule of the graviton propagator can be written as follows:
	\begin{equation} \label{eqn:decomposition_graviton_propagator}
		\begin{split}
			\Phi \left ( \cgreen{p-graviton} \right ) & = - \frac{2 \imaginary p^2}{p^2 + \imaginary \epsilon} \left ( \mathbbit{\mettransstr}_{\mu \nu \rho \sigma} - \left ( 1 - \zeta \right ) \mathbbit{L}_{\mu \nu \rho \sigma} \right ) \\
			& = - \frac{2 \imaginary p^2}{p^2 + \imaginary \epsilon} \left ( \mathbbit{T}_{\mu \nu \rho \sigma} + \zeta \mathbbit{L}_{\mu \nu \rho \sigma} \right )
		\end{split}
	\end{equation}
	Furthermore, the Feynman rules of the graviton propagator and the graviton-ghost propagator are related as follows, where the vector arrows indicate which legs are contracted with the double indices of the tensors \(\gfgr\) and \(\gtgr\):
	\begin{subequations} \label{eqn:relations_graviton_graviton-ghost_propagators}
	\begin{align}
		& \Phi \bigl ( {\scriptstyle \vec{\gfgr}} \cgreen{p-graviton} \bigr ) = p^2 \Phi \bigl ( \cgreen{p-gravitonghost} {\scriptstyle \vec{\gtgr}} \, \bigr ) \, , \\[0.5\baselineskip]
		& \Phi \bigl ( {\scriptstyle \vec{\gfgr}} \cgreen{p-graviton} {\scriptstyle \cevss{\gfgr}} \bigr ) = p^2 \Phi \bigl ( \cgreen{p-gravitonghost} \, \bigr )
		\intertext{and}
		& \Phi \bigl ( {\scriptstyle \mathbbit{L}} \cgreen{p-graviton} {\scriptstyle \mathbbit{L}} \bigr ) = p^2 \Phi \bigl ( {\scriptstyle \cevss{\gtgr}} \cgreen{p-gravitonghost} {\scriptstyle \vec{\gtgr}} \, \bigr ) \, .
	\end{align}
	\end{subequations}
	In particular, the de Donder gauge fixing is the optimal gauge fixing condition for (effective) Quantum General Relativity.
\end{thm}

\begin{proof}
	\eqnref{eqn:decomposition_graviton_propagator} follows from the previous results in this subsection together with the Feynman rule
	\begin{align}
		\begin{split}
		\Phi \left ( \cgreen{p-graviton} \right ) & = - \frac{2 \imaginary}{p^2 + \imaginary \epsilon} \Biggl ( \left ( \eta_{\mu \rho} \eta_{\nu \sigma} + \eta_{\mu \sigma} \eta_{\nu \rho} - \eta_{\mu \nu} \eta_{\rho \sigma} \right ) \\
		& \phantom{= - \frac{2 \imaginary}{p^2 + \imaginary \epsilon} \Biggl [} - \left ( \frac{1 - \zeta}{p^2} \right ) \left ( \eta_{\mu \rho} p_\nu p_\sigma + \eta_{\mu \sigma} p_\nu p_\rho + \eta_{\nu \rho} p_\mu p_\sigma + \eta_{\nu \sigma} p_\mu p_\rho \right ) \Biggr ) \, .
		\end{split}
		\intertext{From this, \eqnsref{eqn:relations_graviton_graviton-ghost_propagators} follows from the previous results in this subsection together with the Feynman rule}
		\Phi \left ( \cgreen{p-gravitonghost} \right ) & = - \frac{2 \imaginary \zeta}{p^2 + \imaginary \epsilon} \eta_{\rho \sigma} \, .
	\end{align}
	The final claim follows then directly from \eqnref{eqn:decomposition_graviton_propagator}.
\end{proof}

\enter

\begin{rem} \label{rem:md_vs_mdd}
	Consider the metric density decomposition of Goldberg \cite{Goldberg} and Capper et al.\ \cite{Capper_Leibbrandt_Ramon-Medrano,Capper_Medrano,Capper_Namazie}, i.e.\
	\begin{equation}
		\boldsymbol{\phi}^{\mu \nu} \coloneq \frac{1}{\gcoupling} \left ( \sqrt{- \dt{g}} g^{\mu \nu} - \eta^{\mu \nu} \right ) \iff \sqrt{- \dt{g}} g^{\mu \nu} \equiv \eta^{\mu \nu} + \gcoupling \boldsymbol{\phi}^{\mu \nu} \, , \label{eqn:metric_density_decomposition}
	\end{equation}
	together with the gauge fixing functional
	\begin{equation}
		\mathscr{C}^\mu \coloneq \partial_\nu \boldsymbol{\phi}^{\mu \nu} \equiv 0 \, . \label{eqn:metric_density_decomposition_gauge_fixing}
	\end{equation}
	Then, the corresponding graviton propagator is given via
	\begin{equation}
		\Phi \left ( \cgreen{p-graviton} \right ) = - \frac{2 \imaginary}{p^2 \left ( p^2 + \imaginary \epsilon \right )} \left ( \mathbbit{T}^{\mu \nu \rho \sigma} + \zeta \mathbbit{L}^{\mu \nu \rho \sigma} \right ) \, ,
	\end{equation}
	i.e.\ the roles of \(\gtgr\) and \(\gfgr\) are reversed compared to the metric decomposition case given in \eqnref{eqn:decomposition_graviton_propagator}, cf.\ \colref{col:l-tensor-gg-ff-qgr}. In particular, the gauge fixing functional \(\mathscr{C}^\mu \left ( \boldsymbol{\phi} \right )\) is the optimal gauge fixing condition for the metric density decomposition, cf.\ \thmref{thm:feynman-rule-graviton-propagator-lt-decomposition}. This is due to the fact that in this case the graviton field \(\boldsymbol{\phi}^{\mu \nu}\) is a tensor density of weight 1, instead of the ordinary tensor \(h_{\mu \nu}\). This will be studied further in \cite{Prinz_10}, cf.\ \cite[Chapter 5.2]{Prinz_PhD}.
\end{rem}

\enter

\begin{thm} \label{thm:three-valent-contraction-identities-qgr}
	The Feynman rule of the three-valent graviton vertex satisfies the following contraction identities:
	{\allowdisplaybreaks
	\begin{subequations}
	\begin{align}
		& \Phi \left ( {\scriptstyle \gtgr} \tcgreen{v-gravitontriple} _{{\scriptstyle \gtgr}}^{{\scriptstyle \gtgr}} \right ) \simeq_\textup{MC} \Phi \left ( {\scriptstyle \mathbbit{L}} \tcgreen{v-gravitontriple} _{{\scriptstyle \mathbbit{L}}}^{{\scriptstyle \mathbbit{L}}} \right ) \simeq_\textup{MC} 0 \label{eqn:contraction_v-triple-graviton} \, , \\
		& \Phi \left ( {\scriptstyle \gtgrred} \tcgreen{v-gravitontriple} _{{\scriptstyle \mathbbit{P}}}^{{\scriptstyle \mathbbit{P}}} \right ) = 0 \label{eqn:contraction_v-triple-graviton-os-tt}
		\intertext{and thus}
		& \Phi \left ( {\scriptstyle \gtgrred} \tcgreen{v-gravitontriple} _{{\scriptstyle \mathbbit{I}}}^{{\scriptstyle \mathbbit{I}}} \right ) \simeq_\textup{OS} \Phi \left ( {\scriptstyle \gtgrred} \tcgreen{v-gravitontriple} _{{\scriptstyle \mathbbit{L}}}^{{\scriptstyle \mathbbit{T}}} \right ) + \Phi \left ( {\scriptstyle \gtgrred} \tcgreen{v-gravitontriple} _{{\scriptstyle \mathbbit{T}}}^{{\scriptstyle \mathbbit{L}}} \right ) \, , \label{eqn:contraction_v-triple-graviton-os-ii}
	\end{align}
	\end{subequations}
	}%
	where \(\mathbbit{P}\) denotes the projector onto the physical degrees of freedom, i.e.\ first the on-shell reduction by setting \(p^2 = 0\), and then the transversal projection via \(\mathbbit{T}\). In addition, \(\simeq_\textup{OS}\) indicates equality on-shell.
\end{thm}

\begin{proof}
All three identities are checked with a FORM program \cite{Vermaseren,Kuipers_Ueda_Vermaseren_Vollinga,Davies_Kaneko_Marinissen_Ueda_Vermaseren} by the author, cf.\ \appref{sec:explicit_feynman_rules_form} for the setup and Feynman rules. In addition, we emphasize the relation between the three identities via the decompositions \(\mathbbit{L}^{\rho \sigma}_{\mu \nu} = \gfgr^{\rho \sigma}_\tau \gtgr_{\mu \nu}^\tau\) of \eqnref{eqn:decomposition_longitudinal_projection_tensor_qgr}, remind the reader of the reduced gauge transformation contraction \(\gtgr_{\mu \nu}^{\prime \kappa} \coloneq p^2 \gtgr_{\mu \nu}^\kappa\) and recall the decomposition \(\mathbbit{I}_{\mu \nu}^{\rho \sigma} = \mathbbit{T}_{\mu \nu}^{\rho \sigma} + \mathbbit{L}_{\mu \nu}^{\rho \sigma}\) of \eqnref{eqn:defn-t-qgr}.
\end{proof}

\enter

\begin{rem} \label{rem:contraction-single-graviton}
	\eqnsaref{eqn:contraction_v-triple-graviton-os-tt}{eqn:contraction_v-triple-graviton-os-ii} imply that the longitudinal projection of a single graviton results in both transversal on-shell cancellations and propagating longitudinal graviton modes. This is similar to Quantum Yang--Mills theory, cf.\ \remref{rem:contraction-single-gluon}.
\end{rem}

\enter

\begin{thm} \label{thm:three-valent-contraction-identities-qgr-with-matter}
	The Feynman rules of the three-valent interactions of gravitons with scalars, spinors, gauge bosons, gauge ghosts and graviton-ghosts satisfy the following on-shell contraction identities, where we have additionally set \(T_\OS \coloneq \OS \circ T\), \(T^\prime \coloneq q^2 T\) and \(\gtymredos \coloneq \OS \circ \gtymred\):
	{\allowdisplaybreaks
	\begin{align}
		& \Phi \left ( {\scriptstyle \gtgrred} \tcgreen{v-gravitonscalartriple}^{{\scriptstyle \OS}}_{{\scriptstyle \OS}} \right ) = 0 \label{eqn:contraction_v-gravitonscalartriple} \\
		& \Phi \left ( {\scriptstyle \gtgrred} \tcgreen{v-gravitonspinortriple}^{{\scriptstyle \OS}}_{{\scriptstyle \OS}} \right ) = 0 \label{eqn:contraction_v-gravitonspinortriple} \\
		& \hphantom{\Phi \left ( {\scriptstyle \gtgrred} \tcgreen{v-gravitongluontriple}^{{\scriptstyle \OS}}_{{\scriptstyle \OS}} \rule{0pt}{5ex} \right ) = \left ( 1 + \frac{1}{\xi} \right ) \Phi \left ( {\scriptstyle \gtgrred} \tcgreen{v-gravitongluontriple}^{{\scriptstyle T_\OS}}_{{\scriptstyle T_\OS}} \rule{0pt}{5ex} \right )} \notag
	\end{align}
	}
	{\allowdisplaybreaks
	\begin{subequations} \label{eqns:contraction_v-gravitongluontriple}
	\begin{align}
		& \Phi \left ( {\scriptstyle \gtgrred} \tcgreen{v-gravitongluontriple}^{{\scriptstyle \OS}}_{{\scriptstyle \OS}} \rule{0pt}{5ex} \right ) = \left ( 1 + \frac{1}{\xi} \right ) \Phi \left ( {\scriptstyle \gtgrred} \tcgreen{v-gravitongluontriple}^{{\scriptstyle T_\OS}}_{{\scriptstyle T_\OS}} \rule{0pt}{5ex} \right ) \label{eqn:contraction_v-gravitongluontriple-os} \\
		& \Phi \left ( {\scriptstyle \gtgrred} \tcgreen{v-gravitongluontriple}^{{\scriptstyle P}}_{{\scriptstyle P}} \rule{0pt}{5ex} \right ) = 0 \label{eqn:contraction_v-gravitongluontriple-phys} \\
		& \Phi \left ( {\scriptstyle \gtgrred} \tcgreen{v-gravitongluontriple}^{{\scriptstyle \OS}}_{{\scriptstyle \gtymredos}} \rule{0pt}{5ex} \right ) = \Phi \left ( {\scriptstyle \gtgrred} \tcgreen{v-gravitongluontriple}^{{\scriptstyle \gtymredos}}_{{\scriptstyle \OS}} \rule{0pt}{5ex} \right ) = 0 \label{eqn:contraction_v-gravitongluontriple-long-os} \\
		& \Phi \left ( {\scriptstyle \mathbbit{I}} \tcgreen{v-gravitongluontriple}^{{\scriptstyle P}}_{{\scriptstyle \gtymredos}} \rule{0pt}{5ex} \right ) = \Phi \left ( {\scriptstyle \mathbbit{I}} \tcgreen{v-gravitongluontriple}^{{\scriptstyle \gtymredos}}_{{\scriptstyle P}} \rule{0pt}{5ex} \right ) = 0 \label{eqn:contraction_v-gravitongluontriple-gluonlongosgluonP} \\
		& \Phi \left ( {\scriptstyle \mathbbit{I}} \tcgreen{v-gravitongluontriple}^{{\scriptstyle T^\prime}}_{{\scriptstyle \gtymredos}} \rule{0pt}{5ex} \right ) = \Phi \left ( {\scriptstyle \mathbbit{I}} \tcgreen{v-gravitongluontriple}^{{\scriptstyle \gtymredos}}_{{\scriptstyle T^\prime}} \rule{0pt}{5ex} \right ) = 0 \label{eqn:contraction_v-gravitongluontriple-gluontransgluonlongos} \\
		& \Phi \left ( {\scriptstyle \mathbbit{I}} \tcgreen{v-gravitongluontriple}^{{\scriptstyle \gtymredos}}_{{\scriptstyle \gtymredos}} \rule{0pt}{5ex} \right ) = 0 \label{eqn:contraction_v-gravitongluontriple-gluonslongos}
	\end{align}
	\end{subequations}
	}
	{\allowdisplaybreaks
	\begin{align}
		& \Phi \left ( {\scriptstyle \gtgrred} \tcgreen{v-gravitongluonghosttriple}^{{\scriptstyle \OS}}_{{\scriptstyle \OS}} \right ) = 0 \label{eqn:contraction_v-gravitongluonghosttriple} \\
		& \Phi \left ( {\scriptstyle \gtgrred} \tcgreen{v-gravitonghosttriple}^{{\scriptstyle \OS}}_{{\scriptstyle \OS}} \rule{0pt}{5ex} \right ) - \Phi \left ( {\scriptstyle \OS} \tcgreen{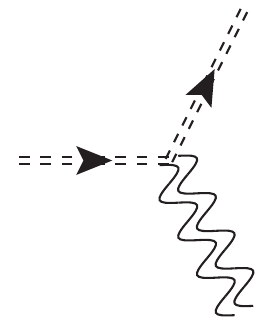}^{{\scriptstyle \OS}}_{\scriptstyle \gtgrred} \right ) = 0 \, , \label{eqn:contraction_v-gravitonghosttriple} \\
		& \hphantom{\Phi \left ( {\scriptstyle \gtgrred} \tcgreen{v-gravitongluontriple}^{{\scriptstyle \OS}}_{{\scriptstyle \OS}} \rule{0pt}{5ex} \right ) = \left ( 1 + \frac{1}{\xi} \right ) \Phi \left ( {\scriptstyle \gtgrred} \tcgreen{v-gravitongluontriple}^{{\scriptstyle T_\OS}}_{{\scriptstyle T_\OS}} \rule{0pt}{5ex} \right )} \notag
	\end{align}
	}%
	and with the corresponding off-shell variants leading to the usual propagator cancellations, as can be seen in the calculations of the proof.
\end{thm}

\begin{proof}
	Again, we consider all momenta incoming and denote the graviton momentum by \(p^\sigma\) and the matter momenta by \(q_1^\sigma\) and \(q_2^\sigma\). Also, we denote the graviton Lorentz indices by \(\mu\) and \(\nu\), respectively. We only calculate the first identity of \eqnref{eqn:contraction_v-gravitongluontriple-long-os}, since the second one follows simply by interchanging the numbers \(1 \! \leftrightarrow \! 2\). In \eqnssaref{eqn:proof_contraction_v-gravitonspinortriple}{eqn:proof_contraction_v-gravitongluonghosttriple}{eqn:proof_contraction_v-gravitonghosttriple} the number 1 corresponds to the particle and the number 2 to the anti-particle. In particular, in \eqnref{eqn:proof_contraction_v-gravitonspinortriple} this implies that the equations of motion differ in a relative sign. In addition, in \eqnref{eqn:proof_contraction_v-gravitongluonghosttriple} we denote the gauge ghost color indices by \(c_1\) and \(c_2\), respectively, and in \eqnref{eqn:proof_contraction_v-gravitonghosttriple} we denote the graviton-ghost Lorentz indices by \(\rho_1\) and \(\rho_2\), respectively. Additionally, in \eqnref{eqn:proof_contraction_v-gravitonghosttriple} we use the symmetric (hermitian) graviton-ghost Lagrange density of \eqnref{eqn:sym-gf-ghost-gr}. With that, we present the actual calculations (the appearing masses \(m\) correspond to the respective matter particles):
	{\allowdisplaybreaks
	\begin{align}
	\begin{split} \label{eqn:proof_contraction_v-gravitonscalartriple}
		\Phi \left ( {\scriptstyle \gtgrred} \tcgreen{v-gravitonscalartriple} \right ) & = \frac{\imaginary \gcoupling}{2} \delta_{i j} \left ( p_\mu \delta_\nu^\kappa + p_\nu \delta_\mu^\kappa \right ) \biggl ( q_1^\mu q_2^\nu + q_1^\nu q_2^\mu - \eta^{\mu \nu} \left ( q_1 \cdot q_2 + m^2 \right ) \biggr ) \\
		& \simeq_\text{MC} - \imaginary \gcoupling \, \delta_{i j} \biggl ( \left ( q_1^2 - m^2 \right ) q_2^\kappa + \left ( q_2^2 - m^2 \right ) q_1^\kappa \biggr ) \\
		& \simeq_\text{OS} 0 \vphantom{\bigg \vert}
		\end{split} \\
		\begin{split} \label{eqn:proof_contraction_v-gravitonspinortriple}
		\Phi \left ( {\scriptstyle \gtgrred} \tcgreen{v-gravitonspinortriple} \right ) & = \frac{\gcoupling}{8} \delta_{k l} \left ( p_\mu \delta_\nu^\kappa + p_\nu \delta_\mu^\kappa \right ) \\ & \phantom{=} \times \biggl ( 2 \eta^{\mu \nu} \left ( \slashed{q}_1 - \slashed{q}_2 - 2m \right ) - \left ( q_1 - q_2 \right )_\mu \gamma_\nu - \left ( q_1 - q_2 \right )_\nu \gamma_\mu \biggr ) \\
		& \simeq_\text{MC} \frac{\gcoupling}{4} \delta_{k l} \Biggl ( \left ( \slashed{q}_1 - m \right ) \left ( - 2 \left ( q_1 + q_2 \right )^\kappa + \left ( q_1 - q_2 \right )^\kappa + \left ( \slashed{q}_1 + m \right ) \gamma^\kappa \right ) \\ & \phantom{\simeq_\text{MC} \frac{\gcoupling}{4 p^2} \Biggl (} + \left ( \slashed{q}_2 + m \right ) \left ( 2 \left ( q_1 + q_2 \right )^\kappa + \left ( q_1 - q_2 \right )^\kappa - \left ( \slashed{q}_2 - m \right ) \gamma^\kappa \right ) \Biggr ) \\
		& \simeq_\text{OS} 0 \vphantom{\bigg \vert}
		\end{split} \\
		\begin{split} \label{eqn:proof_contraction_v-gravitongluontriple-transos}
		\Phi \left ( {\scriptstyle \gtgrred} \tcgreen{v-gravitongluontriple}^{{\scriptstyle \OS}}_{{\scriptstyle \OS}} \rule{0pt}{5ex} \right ) & = \Biggl [ \frac{\imaginary \gcoupling}{2} \left ( p_\mu \delta_\nu^\kappa + p_\nu \delta_\mu^\kappa \right ) \\
		& \phantom{= \Biggl [} \times \delta_{a_1 a_2} \biggl ( \left ( q_1 \cdot q_2 \right ) \left ( \eta^{\mu \nu} \eta^{\rho_1 \rho_2} - \eta^{\mu \rho_1} \eta^{\nu \rho_2} - \eta^{\mu \rho_2} \eta^{\nu \rho_1} \right ) \\
		& \phantom{= \delta_{a_1 a_2} \biggl (} - \eta^{\mu \nu} q_1^{\rho_2} q_2^{\rho_1} - \eta^{\rho_1 \rho_2} \left ( q_1^\mu q_2^\nu + q_2^\mu q_1^\nu \right ) \\
		& \phantom{= \delta_{a_1 a_2} \biggl (} + q_1^{\rho_2} \left ( \eta^{\mu \rho_1} q_2^\nu + \eta^{\nu \rho_1} q_2^\mu \right ) + q_2^{\rho_1} \left ( \eta^{\mu \rho_2} q_1^\nu + \eta^{\nu \rho_2} q_1^\mu \right ) \\
		& \phantom{= \delta_{a_1 a_2} \biggl (} + \frac{1}{\xi} \eta^{\mu \nu} \left ( q_1^{\rho_1} q_2^{\rho_2} + q_1^{\rho_1} q_1^{\rho_2} + q_2^{\rho_1} q_2^{\rho_2} \right ) \\
		& \phantom{= \delta_{a_1 a_2} \biggl (} - \frac{1}{\xi} q_1^{\rho_1} \left ( q_1^\nu \eta^{\mu \rho_2} + q_1^\mu \eta^{\nu \rho_2} \right ) - \frac{1}{\xi} q_2^{\rho_2} \left( q_2^\nu \eta^{\mu \rho_1} + q_2^\mu \eta^{\nu \rho_1} \right ) \biggr ) \Biggr ]_{\text{MC} \, \& \, \text{OS}} \\
		& = \imaginary \gcoupling \left ( 1 + \frac{1}{\xi} \right ) \delta_{a_1 a_2} \biggl ( \left ( q_1 \cdot q_2 \right ) \bigl ( \eta^{\kappa \rho_1} q_2^{\rho_2} + \eta^{\kappa \rho_2} q_1^{\rho_1} \bigr ) - q_1^{\kappa} q_2^{\rho_1} q_2^{\rho_2} - q_2^{\kappa} q_1^{\rho_1} q_1^{\rho_2} \biggr ) \vphantom{\Bigg \vert} \\
		& = \left ( 1 + \frac{1}{\xi} \right ) \Biggl [ \frac{\imaginary \gcoupling}{2} \left ( p_\mu \delta_\nu^\kappa + p_\nu \delta_\mu^\kappa \right ) \times T_{\rho_1}^{\sigma_1} \left ( q_1 \right ) \times T_{\rho_2}^{\sigma_2} \left ( q_2 \right ) \\
		& \phantom{= \Biggl [} \times \delta_{a_1 a_2} \biggl ( \left ( q_1 \cdot q_2 \right ) \left ( \eta^{\mu \nu} \eta^{\rho_1 \rho_2} - \eta^{\mu \rho_1} \eta^{\nu \rho_2} - \eta^{\mu \rho_2} \eta^{\nu \rho_1} \right ) \\
		& \phantom{= \delta_{a_1 a_2} \biggl (} - \eta^{\mu \nu} q_1^{\rho_2} q_2^{\rho_1} - \eta^{\rho_1 \rho_2} \left ( q_1^\mu q_2^\nu + q_2^\mu q_1^\nu \right ) \\
		& \phantom{= \delta_{a_1 a_2} \biggl (} + q_1^{\rho_2} \left ( \eta^{\mu \rho_1} q_2^\nu + \eta^{\nu \rho_1} q_2^\mu \right ) + q_2^{\rho_1} \left ( \eta^{\mu \rho_2} q_1^\nu + \eta^{\nu \rho_2} q_1^\mu \right ) \\
		& \phantom{= \delta_{a_1 a_2} \biggl (} + \frac{1}{\xi} \eta^{\mu \nu} \left ( q_1^{\rho_1} q_2^{\rho_2} + q_1^{\rho_1} q_1^{\rho_2} + q_2^{\rho_1} q_2^{\rho_2} \right ) \\
		& \phantom{= \delta_{a_1 a_2} \biggl (} - \frac{1}{\xi} q_1^{\rho_1} \left ( q_1^\nu \eta^{\mu \rho_2} + q_1^\mu \eta^{\nu \rho_2} \right ) - \frac{1}{\xi} q_2^{\rho_2} \left( q_2^\nu \eta^{\mu \rho_1} + q_2^\mu \eta^{\nu \rho_1} \right ) \biggr ) \Biggr ]_{\text{MC} \, \& \, \text{OS}} \\
		& = \left ( 1 + \frac{1}{\xi} \right ) \Phi \left ( {\scriptstyle \gtgrred} \tcgreen{v-gravitongluontriple}^{{\scriptstyle T_\OS}}_{{\scriptstyle T_\OS}} \rule{0pt}{5ex} \right )
		\end{split} \\
		\begin{split} \label{eqn:proof_contraction_v-gravitongluontriple-phys}
		\Phi \left ( {\scriptstyle \gtgr} \tcgreen{v-gravitongluontriple}^{{\scriptstyle P}}_{{\scriptstyle P}} \right ) & = \Biggl [ \frac{\imaginary \gcoupling}{2} \left ( p_\mu \delta_\nu^\kappa + p_\nu \delta_\mu^\kappa \right ) \\
		& \phantom{= \Biggl [} \times \delta_{a_1 a_2} \biggl ( \left ( q_1 \cdot q_2 \right ) \left ( \eta^{\mu \nu} \eta^{\rho_1 \rho_2} - \eta^{\mu \rho_1} \eta^{\nu \rho_2} - \eta^{\mu \rho_2} \eta^{\nu \rho_1} \right ) \\
		& \phantom{= \delta_{a_1 a_2} \biggl (} - \eta^{\mu \nu} q_1^{\rho_2} q_2^{\rho_1} - \eta^{\rho_1 \rho_2} \left ( q_1^\mu q_2^\nu + q_2^\mu q_1^\nu \right ) \\
		& \phantom{= \delta_{a_1 a_2} \biggl (} + q_1^{\rho_2} \left ( \eta^{\mu \rho_1} q_2^\nu + \eta^{\nu \rho_1} q_2^\mu \right ) + q_2^{\rho_1} \left ( \eta^{\mu \rho_2} q_1^\nu + \eta^{\nu \rho_2} q_1^\mu \right ) \\
		& \phantom{= \delta_{a_1 a_2} \biggl (} + \frac{1}{\xi} \eta^{\mu \nu} \left ( q_1^{\rho_1} q_2^{\rho_2} + q_1^{\rho_1} q_1^{\rho_2} + q_2^{\rho_1} q_2^{\rho_2} \right ) \\
		& \phantom{= \delta_{a_1 a_2} \biggl (} - \frac{1}{\xi} q_1^{\rho_1} \left ( q_1^\nu \eta^{\mu \rho_2} + q_1^\mu \eta^{\nu \rho_2} \right ) - \frac{1}{\xi} q_2^{\rho_2} \left( q_2^\nu \eta^{\mu \rho_1} + q_2^\mu \eta^{\nu \rho_1} \right ) \biggr ) \Biggr ]_{\text{MC} \, \& \, \text{OS}} \\
		& \phantom{=} \times T_{\rho_1}^{\sigma_1} \left ( q_1 \right ) \times T_{\rho_2}^{\sigma_2} \left ( q_2 \right ) \vphantom{\Bigg \vert} \\
		& = \imaginary \gcoupling \left ( 1 + \frac{1}{\xi} \right ) \delta_{a_1 a_2} \biggl ( \left ( q_1 \cdot q_2 \right ) \bigl ( \eta^{\kappa \rho_1} q_2^{\rho_2} + \eta^{\kappa \rho_2} q_1^{\rho_1} \bigr ) - q_1^{\kappa} q_2^{\rho_1} q_2^{\rho_2} - q_2^{\kappa} q_1^{\rho_1} q_1^{\rho_2} \biggr ) \vphantom{\Bigg \vert} \\
		& \phantom{=} \times T_{\rho_1}^{\sigma_1} \left ( q_1 \right ) \times T_{\rho_2}^{\sigma_2} \left ( q_2 \right ) \vphantom{\Bigg \vert} \\
		& = 0 \vphantom{\bigg \vert}
		\end{split} \\
		\begin{split} \label{eqn:proof_contraction_v-gravitongluontriple-long-os}
		\Phi \left ( {\scriptstyle \gtgrred} \tcgreen{v-gravitongluontriple}^{{\scriptstyle I}}_{{\scriptstyle \gtymred}} \rule{0pt}{5ex} \right ) & = \frac{\imaginary \gcoupling}{2} \left ( p_\mu \delta_\nu^\kappa + p_\nu \delta_\mu^\kappa \right ) {q_1}_{\rho_1} \\
		& \phantom{=} \times \delta_{a_1 a_2} \Biggl ( \left ( q_1 \cdot q_2 \right ) \left ( \eta^{\mu \nu} \eta^{\rho_1 \rho_2} - \eta^{\mu \rho_1} \eta^{\nu \rho_2} - \eta^{\mu \rho_2} \eta^{\nu \rho_1} \right ) \\
		& \phantom{= \times \delta_{a_1 a_2} \biggl (} - \eta^{\mu \nu} q_1^{\rho_2} q_2^{\rho_1} - \eta^{\rho_1 \rho_2} \left ( q_1^\mu q_2^\nu + q_2^\mu q_1^\nu \right ) \\
		& \phantom{= \times \delta_{a_1 a_2} \biggl (} + q_1^{\rho_2} \left ( \eta^{\mu \rho_1} q_2^\nu + \eta^{\nu \rho_1} q_2^\mu \right ) + q_2^{\rho_1} \left ( \eta^{\mu \rho_2} q_1^\nu + \eta^{\nu \rho_2} q_1^\mu \right ) \\
		& \phantom{= \times \delta_{a_1 a_2} \biggl (} + \frac{1}{\xi} \eta^{\mu \nu} \left ( q_1^{\rho_1} q_2^{\rho_2} + q_1^{\rho_1} q_1^{\rho_2} + q_2^{\rho_1} q_2^{\rho_2} \right ) \\
		& \phantom{= \times \delta_{a_1 a_2} \biggl (} - \frac{1}{\xi} q_1^{\rho_1} \left ( q_1^\nu \eta^{\mu \rho_2} + q_1^\mu \eta^{\nu \rho_2} \right ) - \frac{1}{\xi} q_2^{\rho_2} \left( q_2^\nu \eta^{\mu \rho_1} + q_2^\mu \eta^{\nu \rho_1} \right ) \Biggr ) \\
		& \simeq_\text{MC} \frac{\imaginary \gcoupling}{\xi} \delta_{a_1 a_2} \biggl ( \eta^{\kappa \rho_2} \left ( q_1^4 + \left ( q_1 \cdot q_2 \right ) q_1^2 \right ) + q_2^2 q_1^\kappa q_2^{\rho_2} - q_1^2 q_2^\kappa q_1^{\rho_2} \biggr ) \\
		& \simeq_\text{OS} 0 \vphantom{\bigg \vert}
		\end{split} \\
		\begin{split}
			\Phi \left ( {\scriptstyle \mathbbit{I}} \tcgreen{v-gravitongluontriple}^{{\scriptstyle P}}_{{\scriptstyle \gtymredos}} \rule{0pt}{5ex} \right ) & = \Biggl [ \imaginary \gcoupling \, {q_1}_{\rho_1} \times \delta_{a_1 a_2} \biggl ( \left ( q_1 \cdot q_2 \right ) \left ( \eta^{\mu \nu} \eta^{\rho_1 \rho_2} - \eta^{\mu \rho_1} \eta^{\nu \rho_2} - \eta^{\mu \rho_2} \eta^{\nu \rho_1} \right ) \\
		& \phantom{= \delta_{a_1 a_2} \biggl (} - \eta^{\mu \nu} q_1^{\rho_2} q_2^{\rho_1} - \eta^{\rho_1 \rho_2} \left ( q_1^\mu q_2^\nu + q_2^\mu q_1^\nu \right ) \\
		& \phantom{= \delta_{a_1 a_2} \biggl (} + q_1^{\rho_2} \left ( \eta^{\mu \rho_1} q_2^\nu + \eta^{\nu \rho_1} q_2^\mu \right ) + q_2^{\rho_1} \left ( \eta^{\mu \rho_2} q_1^\nu + \eta^{\nu \rho_2} q_1^\mu \right ) \\
		& \phantom{= \delta_{a_1 a_2} \biggl (} + \frac{1}{\xi} \eta^{\mu \nu} \left ( q_1^{\rho_1} q_2^{\rho_2} + q_1^{\rho_1} q_1^{\rho_2} + q_2^{\rho_1} q_2^{\rho_2} \right ) \\
		& \phantom{= \delta_{a_1 a_2} \biggl (} - \frac{1}{\xi} q_1^{\rho_1} \left ( q_1^\nu \eta^{\mu \rho_2} + q_1^\mu \eta^{\nu \rho_2} \right ) - \frac{1}{\xi} q_2^{\rho_2} \left( q_2^\nu \eta^{\mu \rho_1} + q_2^\mu \eta^{\nu \rho_1} \right ) \biggr ) \Biggr ]_{\text{MC} \, \& \, \text{OS}} \\
		& \phantom{=} \times T_{\rho_2}^{\sigma_2} \left ( q_2 \right ) \vphantom{\Bigg \vert} \\
		& = \frac{\imaginary \gcoupling}{2 \xi} \, \delta_{a_1 a_2} \, q_2^{\rho_2} \biggl ( \left ( q_1 \cdot q_2 \right ) \eta^{\mu \nu} - q_1^{\mu} q_2^{\nu} - q_1^{\nu} q_2^{\mu} \biggr ) \vphantom{\Bigg \vert} \times T_{\rho_2}^{\sigma_2} \left ( q_2 \right ) \vphantom{\Bigg \vert} \\
		& = 0 \vphantom{\bigg \vert}
		\end{split} \\
		\begin{split} \label{eqn:proof_contraction_v-gravitongluontriple-gravtransgluonslong}
		\Phi \left ( {\scriptstyle \mathbbit{I}} \tcgreen{v-gravitongluontriple}^{{\scriptstyle T^\prime}}_{{\scriptstyle \gtymred}} \rule{0pt}{5ex} \right ) & = \frac{\imaginary \gcoupling}{2} \, {q_1}_{\rho_1} \times q_2^2 \, T_{\rho_2}^{\sigma_2} \left ( q_2 \right ) \\
		& \phantom{=} \times \delta_{a_1 a_2} \Biggl ( \left ( q_1 \cdot q_2 \right ) \left ( \eta^{\mu \nu} \eta^{\rho_1 \rho_2} - \eta^{\mu \rho_1} \eta^{\nu \rho_2} - \eta^{\mu \rho_2} \eta^{\nu \rho_1} \right ) \\
		& \phantom{= \times \delta_{a_1 a_2} \biggl (} - \eta^{\mu \nu} q_1^{\rho_2} q_2^{\rho_1} - \eta^{\rho_1 \rho_2} \left ( q_1^\mu q_2^\nu + q_2^\mu q_1^\nu \right ) \\
		& \phantom{= \times \delta_{a_1 a_2} \biggl (} + q_1^{\rho_2} \left ( \eta^{\mu \rho_1} q_2^\nu + \eta^{\nu \rho_1} q_2^\mu \right ) + q_2^{\rho_1} \left ( \eta^{\mu \rho_2} q_1^\nu + \eta^{\nu \rho_2} q_1^\mu \right ) \\
		& \phantom{= \times \delta_{a_1 a_2} \biggl (} + \frac{1}{\xi} \eta^{\mu \nu} \left ( q_1^{\rho_1} q_2^{\rho_2} + q_1^{\rho_1} q_1^{\rho_2} + q_2^{\rho_1} q_2^{\rho_2} \right ) \\
		& \phantom{= \times \delta_{a_1 a_2} \biggl (} - \frac{1}{\xi} q_1^{\rho_1} \left ( q_1^\nu \eta^{\mu \rho_2} + q_1^\mu \eta^{\nu \rho_2} \right ) - \frac{1}{\xi} q_2^{\rho_2} \left( q_2^\nu \eta^{\mu \rho_1} + q_2^\mu \eta^{\nu \rho_1} \right ) \Biggr ) \\
		& \simeq_\text{MC} \frac{\imaginary \gcoupling}{2 \xi} \delta_{a_1 a_2} \, q_1^2 \Biggl ( q_2^2 \Bigl ( \eta^{\mu \nu} q_1^{\sigma_2} - \eta^{\mu \sigma_2} q_1^\nu - \eta^{\nu \sigma_2} q_1^\mu \Bigr ) \\
		& \phantom{\simeq_\text{MC} \frac{\imaginary \gcoupling}{2 \xi} \delta_{a_1 a_2} \, q_1^2 \Biggl (} + q_2^{\sigma_2} \Bigl ( q_1^\mu q_2^\nu + q_1^\nu q_2^\mu - \left ( q_1 \cdot q_2 \right ) \eta^{\mu \nu} \Bigr ) \Biggr ) \\
		& \simeq_\text{OS-1} 0 
		\end{split} \\
		\begin{split} \label{eqn:proof_contraction_v-gravitongluontriple-gravgluonslong}
		\Phi \left ( {\scriptstyle \mathbbit{I}} \tcgreen{v-gravitongluontriple}^{{\scriptstyle \gtymred}}_{{\scriptstyle \gtymred}} \rule{0pt}{5ex} \right ) & = \frac{\imaginary \gcoupling}{2} {q_1}_{\rho_1} {q_2}_{\rho_2} \\
		& \phantom{=} \times \delta_{a_1 a_2} \Biggl ( \left ( q_1 \cdot q_2 \right ) \left ( \eta^{\mu \nu} \eta^{\rho_1 \rho_2} - \eta^{\mu \rho_1} \eta^{\nu \rho_2} - \eta^{\mu \rho_2} \eta^{\nu \rho_1} \right ) \\
		& \phantom{= \times \delta_{a_1 a_2} \biggl (} - \eta^{\mu \nu} q_1^{\rho_2} q_2^{\rho_1} - \eta^{\rho_1 \rho_2} \left ( q_1^\mu q_2^\nu + q_2^\mu q_1^\nu \right ) \\
		& \phantom{= \times \delta_{a_1 a_2} \biggl (} + q_1^{\rho_2} \left ( \eta^{\mu \rho_1} q_2^\nu + \eta^{\nu \rho_1} q_2^\mu \right ) + q_2^{\rho_1} \left ( \eta^{\mu \rho_2} q_1^\nu + \eta^{\nu \rho_2} q_1^\mu \right ) \\
		& \phantom{= \times \delta_{a_1 a_2} \biggl (} + \frac{1}{\xi} \eta^{\mu \nu} \left ( q_1^{\rho_1} q_2^{\rho_2} + q_1^{\rho_1} q_1^{\rho_2} + q_2^{\rho_1} q_2^{\rho_2} \right ) \\
		& \phantom{= \times \delta_{a_1 a_2} \biggl (} - \frac{1}{\xi} q_1^{\rho_1} \left ( q_1^\nu \eta^{\mu \rho_2} + q_1^\mu \eta^{\nu \rho_2} \right ) - \frac{1}{\xi} q_2^{\rho_2} \left( q_2^\nu \eta^{\mu \rho_1} + q_2^\mu \eta^{\nu \rho_1} \right ) \Biggr ) \\
		& \simeq_\text{MC} \frac{\imaginary \gcoupling}{2 \xi} \delta_{a_1 a_2} \biggl ( \eta^{\mu \nu} \Bigl ( \left ( q_1 \cdot q_2 \right ) \bigl ( q_1^2 + q_2^2 \bigr ) + q_1^2 q_2^2 \Bigr ) - \bigl ( q_1^\mu q_2^\nu + q_1^\nu q_2^\mu \bigr ) \bigl ( q_1^2 + q_2^2 \bigr ) \biggr ) \\
		& \simeq_\text{OS} 0 \vphantom{\bigg \vert}
		\end{split} \\
		\begin{split} \label{eqn:proof_contraction_v-gravitongluonghosttriple}
		\Phi \left ( {\scriptstyle \gtgrred} \tcgreen{v-gravitongluonghosttriple} \right ) & = \frac{\imaginary \gcoupling}{2 \xi} \left ( p_\mu \delta_\nu^\kappa + p_\nu \delta_\mu^\kappa \right ) \delta_{a b} \, \biggl ( \left ( q_1 \cdot q_2 \right ) \eta^{\mu \nu} - q_1^\mu q_2^\nu - q_2^\mu q_1^\nu \biggr ) \\
		& \simeq_\text{MC} \frac{\imaginary \gcoupling}{\xi} \delta_{a b} \left ( q_2^2 q_1^\kappa + q_1^2 q_2^\kappa \right ) \\
		& \simeq_\text{OS} 0 \vphantom{\bigg \vert}
		\end{split} \\
		\begin{split} \label{eqn:proof_contraction_v-gravitonghosttriple}
		\Phi \left ( {\scriptstyle \gtgrred} \tcgreen{v-gravitonghosttriple} \right ) & = \frac{\imaginary \gcoupling}{8} \left ( p_\mu \delta_\nu^\kappa + p_\nu \delta_\mu^\kappa \right ) \\
		& \phantom{=} \times \Biggl ( \Bigl ( q_2^\mu \bigl ( q_2^{\sigma_1} + 2 q_1^{\sigma_1} \bigr ) - q_1^\mu q_2^{\sigma_1} \Bigr ) \eta^{\nu \sigma_2} + \Bigl ( q_1^\mu \bigl ( q_1^{\sigma_2} + 2 q_2^{\sigma_2} \bigr ) - q_2^\mu q_1^{\sigma_2} \Bigr ) \eta^{\nu \sigma_1} \\
		& \hphantom{= \times} \mkern-5mu + \Bigl ( q_2^\nu \bigl ( q_2^{\sigma_1} + 2 q_1^{\sigma_1} \bigr ) - q_1^\nu q_2^{\sigma_1} \Bigr ) \eta^{\mu \sigma_2} + \Bigl ( q_1^\nu \bigl ( q_1^{\sigma_2} + 2 q_2^{\sigma_2} \bigr ) - q_2^\nu q_1^{\sigma_2} \Bigr ) \eta^{\mu \sigma_1} \\
		& \hphantom{= \times} \mkern-5mu - \bigl ( q_1^{\sigma_1} q_1^{\sigma_2} + 2 q_1^{\sigma_1} q_2^{\sigma_2} + q_2^{\sigma_1} q_2^{\sigma_2} \bigr ) \eta^{\mu \nu} - \left ( q_1 + q_2 \right )^2 \bigl ( \eta^{\mu \sigma_1} \eta^{\nu \sigma_2} + \eta^{\mu \sigma_2} \eta^{\nu \sigma_1} \bigr ) \Biggr ) \\
		& \simeq_\text{MC} \frac{\imaginary \gcoupling}{4} \Biggl ( \Big ( 2 q_2^{\sigma_1} \big ( q_1^2 + q_1 \cdot q_2 \big ) + q_1^{\sigma_1} \big ( q_1^2 - q_2^2 \big ) \Big ) \eta^{\sigma_2 \kappa} \\
		& \hphantom{\simeq_\text{MC} \frac{\imaginary \gcoupling}{4 p^2} \Biggl (} + \Big ( 2 q_1^{\sigma_2} \big ( q_1 \cdot q_2 + q_2^2 \big ) - q_2^{\sigma_2} \big ( q_1^2 - q_2^2 \big ) \Big ) \eta^{\sigma_1 \kappa} \Biggr ) \\
		& \simeq_\text{OS} \frac{\imaginary \gcoupling}{2} \left ( q_1 \cdot q_2 \right ) \Bigl ( q_1^{\sigma_2} \eta^{\sigma_1 \kappa} + q_2^{\sigma_1} \eta^{\sigma_2 \kappa} \Bigr ) \\
		& \simeq_{\text{MC} \, \& \, \text{OS}} \Phi \left ( \tcgreen{v-gravitonghosttriple-rot}_{\scriptstyle \gtgrred} \right ) \, ,
		\end{split}
	\end{align}
	}%
	where \(\simeq_\text{MC}\) indicates equality modulo momentum conservation, i.e.\ setting \(p^\sigma \coloneq - q_1^\sigma - q_2^\sigma\), \(\simeq_\text{OS}\) indicates equality on-shell, i.e.\ setting \(q_1^2 = q_2^2 = 0\), and \(\simeq_\text{OS-1}\) indicates equality with simply particle 1 on-shell, i.e.\ setting \(q_1^2 = 0\). In addition, \(\simeq_{\text{MC} \, \& \, \text{OS}}\) indicates equality modulo first momentum conservation and then on-shell reduction.
\end{proof}

\enter

\begin{col} \label{col:three-valent-contraction-identities-gravitongluontriple}
	Given the situation of \thmref{thm:three-valent-contraction-identities-qgr-with-matter}, we obtain the following special cases for the graviton--two-gauge-boson vertex, where we have additionally set \(\mathbbit{T}^\prime \coloneq p^2 \mathbbit{T}\):
	\begin{subequations} \label{eqns:contraction_v-gravitongluontriple_sc}
	\begin{align}
		& \Phi \left ( {\scriptstyle \gtgrred} \tcgreen{v-gravitongluontriple}^{{\scriptstyle P}}_{{\scriptstyle \gtymredos}} \rule{0pt}{5ex} \right ) = \Phi \left ( {\scriptstyle \gtgrred} \tcgreen{v-gravitongluontriple}^{{\scriptstyle \gtymredos}}_{{\scriptstyle P}} \rule{0pt}{5ex} \right ) = 0 \label{eqn:contraction_v-gravitongluontriple-longosphys} \\
		& \Phi \left ( {\scriptstyle \gtgrred} \tcgreen{v-gravitongluontriple}^{{\scriptstyle T^\prime}}_{{\scriptstyle \gtymredos}} \rule{0pt}{5ex} \right ) = \Phi \left ( {\scriptstyle \gtgrred} \tcgreen{v-gravitongluontriple}^{{\scriptstyle \gtymredos}}_{{\scriptstyle T^\prime}} \rule{0pt}{5ex} \right ) = 0 \label{eqn:contraction_v-gravitongluontriple-longostransred} \\
		& \Phi \left ( {\scriptstyle \gtgrred} \tcgreen{v-gravitongluontriple}^{{\scriptstyle \gtymredos}}_{{\scriptstyle \gtymredos}} \rule{0pt}{5ex} \right ) = 0 \label{eqn:contraction_v-gravitongluontriple-long} \\
		& \Phi \left ( {\scriptstyle \mathbbit{T}^\prime} \tcgreen{v-gravitongluontriple}^{{\scriptstyle P}}_{{\scriptstyle \gtymredos}} \rule{0pt}{5ex} \right ) = \Phi \left ( {\scriptstyle \mathbbit{T}^\prime} \tcgreen{v-gravitongluontriple}^{{\scriptstyle \gtymredos}}_{{\scriptstyle P}} \rule{0pt}{5ex} \right ) = 0 \label{eqn:transred_v-gravitongluontriple-longosphys} \\
		& \Phi \left ( {\scriptstyle \mathbbit{T}^\prime} \tcgreen{v-gravitongluontriple}^{{\scriptstyle T^\prime}}_{{\scriptstyle \gtymredos}} \rule{0pt}{5ex} \right ) = \Phi \left ( {\scriptstyle \mathbbit{T}^\prime} \tcgreen{v-gravitongluontriple}^{{\scriptstyle \gtymredos}}_{{\scriptstyle T^\prime}} \rule{0pt}{5ex} \right ) = 0 \label{eqn:transred_v-gravitongluontriple-longostransred} \\
		& \Phi \left ( {\scriptstyle \mathbbit{T}^\prime} \tcgreen{v-gravitongluontriple}^{{\scriptstyle \gtymredos}}_{{\scriptstyle \gtymredos}} \rule{0pt}{5ex} \right ) = 0 \label{eqn:transred_v-gravitongluontriple-long}
	\end{align}
	\end{subequations}
	and with the corresponding off-shell variants leading to the usual propagator cancellations, as can be seen in the calculations of the proof of \thmref{thm:three-valent-contraction-identities-qgr-with-matter}.
\end{col}

\begin{proof}
	All six identities follow directly from \eqnssaref{eqn:contraction_v-gravitongluontriple-gluonlongosgluonP}{eqn:contraction_v-gravitongluontriple-gluontransgluonlongos}{eqn:contraction_v-gravitongluontriple-gluonslongos} by replacing \(\mathbbit{I}\) via \(\gtgrred\) and \(\mathbbit{T}^\prime\).
\end{proof}

\enter

\begin{rem} \label{rem:three-valent-contraction-identities-qgr-with-matter}
	We emphasize that the graviton--two-gauge-boson vertex Feynman rule of \eqnref{eqn:fr-v-gravitongluontriple} does not vanish under any combination of longitudinal and transversal projection operators, i.e.\ operators from the set \(\mathcal{T}_\text{QGR-SM} \equiv \set{L, I, T} \cup \set{\mathbbit{L}, \mathbbit{I}, \mathbbit{T}}\). This is unlike the situation of the triple gauge boson and triple graviton vertex Feynman rules of \eqnsaref{eqn:fr-v-gluontriple}{eqn:fr-v-gravitontriple}, cf.\ \thmsaref{thm:three-valent-contraction-identities-qym}{thm:three-valent-contraction-identities-qgr} and \remsaref{rem:contraction-single-gluon}{rem:contraction-single-graviton}. However, if we additionally consider on-shell reductions of one or both gauge bosons, we find many independent relations in \eqnsref{eqns:contraction_v-gravitongluontriple} of \thmref{thm:three-valent-contraction-identities-qgr-with-matter} and some notable special cases in \eqnsref{eqns:contraction_v-gravitongluontriple_sc} of \colref{col:three-valent-contraction-identities-gravitongluontriple}. In particular, our results can be summarized in the following table, where ``\(\checkmark\)'' denotes a vanishing contraction identity and ``\(\times\)'' a non-vanishing contraction identity, respectively:
	\begin{center}
	\renewcommand{\arraystretch}{1.25}
	\begin{tabular}{c|cccccc}
		& \(\left ( \OS, \OS \right )\)
		& \(\left ( P, P \right )\)
		& \(\left ( \gtymredos, \OS \right )\)
		& \(\left ( \gtymredos, P \right )\)
		& \(\left ( \gtymredos, T^\prime \right )\)
		& \(\left ( \gtymredos, \gtymredos \right )\) \\
		\hline
		\(\gtgrred\) & \(\times\) & \(\checkmark\) & \(\checkmark\) & \(\checkmark\) & \(\checkmark\) & \(\checkmark\) \\
		\(\mathbbit{I}\) & \(\times\) & \(\times\) & \(\times\) & \(\checkmark\) & \(\checkmark\) & \(\checkmark\) \\
		\(\mathbbit{T}^\prime\) & \(\times\) & \(\times\) & \(\times\) & \(\checkmark\) & \(\checkmark\) & \(\checkmark\) \\
	\end{tabular}
	\end{center}
	Here, we have used again the definitions \(\mathbbit{T}^\prime \coloneq p^2 \mathbbit{T}\), \(T_\OS \coloneq \OS \circ T\), \(T^\prime \coloneq q^2 T\) and \(\gtymredos \coloneq \OS \circ \gtymred\). Additionally, we emphasize the following concrete result for \eqnref{eqn:contraction_v-gravitongluontriple-long} before the on-shell limit
	\begin{align}
		\begin{split}
		\Phi \left ( {\scriptstyle \gtgrred} \tcgreen{v-gravitongluontriple}^{{\scriptstyle \gtymred}}_{{\scriptstyle \gtymred}} \right ) & \simeq_\text{MC} \frac{\imaginary \gcoupling}{\xi} \delta_{a_1 a_2} \left ( q_2^4 q_1^\kappa + q_1^4 q_2^\kappa \right ) \\
		& \simeq_\text{OS} 0 \, , \vphantom{\bigg \vert}
		\end{split}
		\intertext{which looks structurally very similar to the corresponding gauge ghost identity of \eqnref{eqn:contraction_v-gravitongluonghosttriple}}
		\begin{split}
		\Phi \left ( {\scriptstyle \gtgrred} \tcgreen{v-gravitongluonghosttriple} \right ) & \simeq_\text{MC} \frac{\imaginary \gcoupling}{\xi} \delta_{a b} \left ( q_2^2 q_1^\kappa + q_1^2 q_2^\kappa \right ) \\
		& \simeq_\text{OS} 0 \vphantom{\bigg \vert} \, ,
		\end{split}
	\end{align}
	however, they are not algebraically related due to different powers in their momenta. Finally, we remark that the longitudinal projection of the graviton--graviton-ghost--graviton-antighost vertex vanishes on-shell against the sign-permuted on-shell variant of itself with the longitudinally projected graviton exchanged with the graviton-ghost, where the minus sign is due to the odd parity of ghosts and gaugeons, cf.\ \eqnsaref{eqn:contraction_v-gravitonghosttriple}{eqn:proof_contraction_v-gravitonghosttriple}: Notably, these are the usual Faddeev--Popov cancellation identities, which simplify in the symmetric Yang--Mills case to the single contraction given in \eqnsaref{eqn:contraction_v-gluonghosttriple}{eqn:proof_contraction_v-gluonghosttriple}.
\end{rem}

\section{Conclusion} \label{sec:conclusion}

We have studied the transversal structure of (effective) Quantum General Relativity coupled to the Standard Model. To this end, we provided the corresponding propagator and three-valent Feynman rules in \sectionref{sec:explicit_feynman_rules}. Then we discussed several aspects of the corresponding longitudinal, identical and transversal projection tensors in \sectionref{sec:longitudinal_and_transversal_projections}. In particular, we recalled known and immediate identities of Quantum Yang--Mills theory in \ssecref{ssec:transversality_qym} and then proceeded by analogy to introduce their novel and involved counterparts in (effective) Quantum General Relativity in \ssecref{ssec:transversality_qgr}. Our main results are the following: First we discussed the decomposition of the gauge boson and graviton propagators into their physical and unphysical degrees of freedom in \thmsaref{thm:feynman-rule-gluon-propagator-lt-decomposition}{thm:feynman-rule-graviton-propagator-lt-decomposition}. Next we studied the corresponding cancellation identities for the pure theories in \thmsaref{thm:three-valent-contraction-identities-qym}{thm:three-valent-contraction-identities-qgr} and then ultimately for their couplings to matter from the Standard Model in \thmsaref{thm:three-valent-contraction-identities-qym-with-matter}{thm:three-valent-contraction-identities-qgr-with-matter}. We believe that these results provide further insight into the involved tensorial structure of gravitational Feynman integrals. Notably, we also provide the derived QGR-SM Feynman rules in FORM code \cite{Vermaseren,Kuipers_Ueda_Vermaseren_Vollinga,Davies_Kaneko_Marinissen_Ueda_Vermaseren} in \appendixref{sec:explicit_feynman_rules_form} and via the supplementary file \texttt{QGR-SM-FR.frm}, available on the journal homepage and the arXiv preprint server. This continues the aim to prove the renormalizability of (effective) Quantum General Relativity (QGR), possibly coupled to matter from the Standard Model (SM), as was suggested in \cite{Kreimer_QG1} and then worked out in \cite{Prinz_3}. In particular, we refer to \cite{Prinz_2,Prinz_4} for introductions to the corresponding perturbative expansions. Additionally, the present work also relates to the BRST double complex of diffeomorphisms and gauge transformations \cite{Prinz_5} and the respective symmetric (hermitian) ghost Lagrange densities \cite{Prinz_6}. Also, we emphasize the corresponding discussion for a non-vanishing cosmological constant \(\Lambda\) in \cite{Prinz_8}. Finally, we aim to combine these different angles on the renormalization problem of QGR-SM in \cite{Prinz_9} with the introduction of a differential-graded renormalization Hopf algebra together with an appropriate UV completion thereof. This article is based on the author's dissertation \cite{Prinz_PhD}.

\section*{Acknowledgments}
\addcontentsline{toc}{section}{Acknowledgments}

The author thanks Christian Blohmann and Albrecht Klemm for general feedback and supportive mentoring. This research is supported by the \emph{Kolleg Mathematik Physik Berlin} of the Humboldt University of Berlin and the University of Potsdam via the research group of Sylvie Paycha.

\begin{appendix}

\section{Explicit Feynman rules in FORM code} \label{sec:explicit_feynman_rules_form}

Finally, we present the Feynman rules from \sectionref{sec:explicit_feynman_rules} in FORM code \cite{Vermaseren,Kuipers_Ueda_Vermaseren_Vollinga,Davies_Kaneko_Marinissen_Ueda_Vermaseren}, so that they can be used for computer assisted calculations. Additionally, we provide them also in the separate file \texttt{QGR-SM-FR.frm} for download in ASCII format, available on the journal homepage and the arXiv preprint server: This avoids any copy and paste error due to the PDF reader. Therein, we have defined the propagators and vertices as functions of their fields, momenta and indices, whereas in the present appendix we use the diagrammatic notation from \sectionref{sec:explicit_feynman_rules}. Also, therein we have spelled out the three-graviton vertex Feynman rule including all its permutations and symmetrizations, while printing the condensed version here for better readability. Again, we acknowledge the use of \texttt{JaxoDraw} for our drawings \cite{Binosi_Theussl,Binosi_Collins_Kaufhold_Theussl}. The following FORM preamble defines and explains all following quantities:

\begin{equation}
	\begin{split}
		& \texttt{* --- Constants / parameters (scalars) ---} \\
		& \texttt{Symbols m, eps, xi, zeta, gcpl, vkappa;} \\
		& \texttt{*   m, eps     : scalar mass, epsilon (i*epsilon prescription)} \\
		& \texttt{*   xi, zeta  : gauge parameters} \\
		& \texttt{*   gcpl, vkappa : coupling constants} \\[\baselineskip]
		& \texttt{* --- Dimensions (scalars) ---} \\
		& \texttt{Symbols SptDim, AdjDim, ScalDim, FermDim;} \\
		& \texttt{Dimension SptDim;      * default = spacetime dimension (applies to vectors)} \\[\baselineskip]
		& \texttt{* --- Momenta (vectors) ---} \\
		& \texttt{Vectors p, p1, p2, p3, q1, q2;} \\[\baselineskip]
		& \texttt{* --- Lorentz indices (dimension SptDim) ---} \\
		& \texttt{Indices mu=SptDim, nu=SptDim, rho=SptDim, sigma=SptDim,} \\
		& \texttt{        r1=SptDim, r2=SptDim, r3=SptDim, s1=SptDim, s2=SptDim;} \\
		& \texttt{Indices m1=SptDim, m2=SptDim, m3=SptDim,} \\
		& \texttt{        n1=SptDim, n2=SptDim, n3=SptDim;} \\[\baselineskip]
		& \texttt{* --- Adjoint color indices (dimension AdjDim) ---} \\
		& \texttt{Indices a=AdjDim, b=AdjDim, c=AdjDim,} \\
		& \texttt{        a1=AdjDim, a2=AdjDim, a3=AdjDim;} \\[\baselineskip]
		& \texttt{* --- Scalar representation / matrix indices (dimension ScalDim) ---} \\
		& \texttt{Indices i=ScalDim, j=ScalDim;} \\[\baselineskip]
		& \texttt{* --- Fermion representation / matrix indices (dimension FermDim) ---} \\
		& \texttt{Indices k=FermDim, l=FermDim;} \\[\baselineskip]
		& \texttt{* --- Color matrices / structure constants (tensors) ---} \\
		& \texttt{Tensor f(antisymmetric);   * structure constants f{\usc}{abc}} \\
		& \texttt{Tensor H;                  * scalar representation matrix (H{\usc}a){\usc}{ij}} \\
		& \texttt{Tensor S;                  * fermion representation matrix (S{\usc}a){\usc}{kl}} \\[\baselineskip]
		& \texttt{* Built in, no declaration needed:} \\
		& \texttt{*   i{\usc}  (imaginary unit),  d{\usc}  (Kronecker delta / metric),} \\
		& \texttt{*   g{\usc}(L,...)  (Dirac matrices),  gi{\usc}(L)  (unit Dirac matrix)}
	\end{split}
\end{equation}

\subsection{QGR-SM propagators} \label{ssec:gravity-matter-propagators_form}

The arrows denote the particle flow (if applicable); all momenta are considered from left to right. We note that our graviton propagator Feynman rule in \eqnref{eqn:fr-p-graviton-form} differs from the main literature, e.g.\ \cite{Choi_Shim_Song}, in that we do not assume the Feynman gauge \(\zeta \coloneq 1\) for the linearized de Donder gauge fixing condition \(\lindeDonder_\mu \coloneq \eta^{\rho \sigma} \Gamma_{\mu \rho \sigma}\) and instead consider a general gauge parameter \(\zeta \in \mathbb{R}\), cf.\ \cite[Theorem 4.12]{Prinz_4}. Additionally, unlike e.g.\ \cite{Romao_Silva}, we use rescaled ghost Lagrange densities such that the gauge ghost and graviton-ghost propagators \eqnsaref{eqn:fr-p-gluonghost-form}{eqn:fr-p-gravitonghost-form} have the same scaling in \(\xi\) and \(\zeta\) as the longitudinal contributions of the corresponding graviton and gauge boson propagators \eqnsaref{eqn:fr-p-gluon-form}{eqn:fr-p-graviton-form}, cf.\ \cite{Prinz_5,Prinz_6} for the corresponding BRST setup:
{\allowdisplaybreaks
\begin{align}
	\Phi \left ( \scriptstyle i \cgreen{p-scalar} j \right ) & = \texttt{i{\usc}/(p.p - m\textasciicircum{}2 + i{\usc}*eps)*d{\usc}(i,j)} \label{eqn:fr-p-scalar-form} \\
		\Phi \left ( \scriptstyle k \cgreen{p-spinor} l \right ) & = \texttt{i{\usc}*(g{\usc}(1,p) + m*gi{\usc}(1))/(p.p - m\textasciicircum{}2 + i{\usc}*eps)*d{\usc}(k,l)} \label{eqn:fr-p-spinor-form} \\
	\begin{split} \label{eqn:fr-p-gluon-form}
		\Phi \left ( \scriptstyle \mu , a \cgreen{p-gluon} \nu , b \right ) & = \texttt{-i{\usc}/(p.p + i{\usc}*eps)*d{\usc}(a,b)} \\
		& \phantom{=} \texttt{*(d{\usc}(mu,nu) - (1-xi)/p.p*p(mu)*p(nu))}
	\end{split} \\
	\Phi \left ( \scriptstyle a \cgreen{p-gluonghost} b \right ) & = \texttt{-i{\usc}*xi/(p.p + i{\usc}*eps)*d{\usc}(a,b)} \label{eqn:fr-p-gluonghost-form} \\
	\begin{split} \label{eqn:fr-p-graviton-form}
		\Phi \left ( \scriptstyle \mu \nu \cgreen{p-graviton} \rho \sigma \right ) & = \texttt{-2*i{\usc}/(p.p + i{\usc}*eps)*(d{\usc}(mu,rho)*d{\usc}(nu,sigma)} \\ & \phantom{=} \texttt{+ d{\usc}(mu,sigma)*d{\usc}(nu,rho) - d{\usc}(mu,nu)*d{\usc}(rho,sigma)} \\ & \phantom{=} \texttt{- (1-zeta)/p.p*(d{\usc}(mu,rho)*p(nu)*p(sigma)} \\ & \phantom{=} \texttt{+ d{\usc}(mu,sigma)*p(nu)*p(rho) + d{\usc}(nu,rho)*p(mu)*p(sigma)} \\ & \phantom{=} \texttt{+ d{\usc}(nu,sigma)*p(mu)*p(rho)))}
	\end{split} \\
	\Phi \left ( \scriptstyle \rho \cgreen{p-gravitonghost} \sigma \right ) & = \texttt{-2*i{\usc}*zeta/(p.p + i{\usc}*eps)*d{\usc}(rho,sigma)} \label{eqn:fr-p-gravitonghost-form}
\end{align}
}%

\subsection{QGR-SM 3-valent vertices} \label{ssec:three-val-gravity-matter-vertices_form}

The arrows denote the particle flow (if applicable); all momenta are considered incoming. In addition, the sum over \(\operatorname{Perm} \left ( 1, 2, 3 \right )\) runs over all permutations of \(\set{1, 2, 3}\) and the sum over \(\mu_i \leftrightarrow \nu_i\) means symmetrization between \(\mu_i\) and \(\nu_i\). We remark that our graviton vertex Feynman rule in \eqnref{eqn:fr-v-gravitontriple-form} differs from the main literature, e.g.\ \cite{Choi_Shim_Song}, in that we do not assume the de Donder relation \(\partial^\nu h_{\mu \nu} = \eta^{\rho \sigma} \partial_\mu h_{\rho \sigma}\) to simplify its derivation, cf.\ \cite[Theorem 4.10]{Prinz_4}. Also, we present the graviton vertex Feynman rule in a condensed expression, while the fully symmetrized expression can be found in the \texttt{QGR-SM-FR.frm} supplementary file, available on the journal homepage and the arXiv preprint server. Additionally, unlike e.g.\ \cite{Romao_Silva}, we also use a general gauge parameter \(\xi \in \mathbb{R}\) for the graviton--gauge-boson vertex in \eqnref{eqn:fr-v-gravitongluontriple-form} and our rescaling convention for the ghost Lagrange densities also affects the graviton--gluon-ghost vertex Feynman rule in \eqnref{eqn:fr-v-gravitongluonghosttriple-form}, cf.\ again \cite{Prinz_5,Prinz_6} for the corresponding BRST setup:
{\allowdisplaybreaks
\begin{align}
	\Phi \left ( \scriptstyle \rho, a \tcgreen{v-gluonscalartriple}^{\scriptstyle q_2, j}_{\scriptstyle q_1, i} \right ) & = \texttt{-gcpl*(q1(rho) - q2(rho))*H(a,i,j)} \label{eqn:fr-v-gluonscalartriple-form} \\
	\Phi \left ( \scriptstyle \rho, a \tcgreen{v-gluonspinortriple}^{\scriptstyle l}_{\scriptstyle k} \right ) & = \texttt{-i{\usc}*gcpl*g{\usc}(1,rho)*S(a,k,l)} \label{eqn:fr-v-gluonspinortriple-form} \\
	\begin{split} \label{eqn:fr-v-gluontriple-form}
		\Phi \left ( \scriptstyle 1 \tcgreen{v-gluontriple}^{\scriptstyle 3}_{\scriptstyle 2} \rule{0pt}{5ex} \right ) & = \texttt{-gcpl*f(a1,a2,a3)*(d{\usc}(r1,r2)*(p1(r3) - p2(r3))} \\
		& \phantom{=} \mkern-22mu \texttt{+ d{\usc}(r2,r3)*(p2(r1) - p3(r1)) + d{\usc}(r3,r1)*(p3(r2) - p1(r2)))}
	\end{split} \\
	\Phi \left ( \scriptstyle \rho , a \tcgreen{v-gluonghosttriple}^{\scriptstyle q_2, c}_{\scriptstyle q_1, b} \right ) & = \texttt{gcpl/2*f(a,b,c)*(q1(rho) - q2(rho))} \label{eqn:fr-v-gluonghosttriple-form} \\
	\begin{split} \label{eqn:fr-v-gravitonscalartriple-form}
		\Phi \left ( \scriptstyle \mu \nu \tcgreen{v-gravitonscalartriple}^{\scriptstyle q_2, j}_{\scriptstyle q_1, i} \right ) & = \texttt{i{\usc}*vkappa/2*d{\usc}(i,j)*(q1(mu)*q2(nu) + q1(nu)*q2(mu)} \\
		& \phantom{=} \texttt{- d{\usc}(mu,nu)*(q1.q2 + m\textasciicircum{}2))}
	\end{split} \\
	\begin{split} \label{eqn:fr-v-gravitonspinortriple-form}
		\Phi \left ( \scriptstyle \mu \nu \tcgreen{v-gravitonspinortriple}^{\scriptstyle q_2, l}_{\scriptstyle q_1, k} \right ) & = \texttt{vkappa/8*d{\usc}(k,l)*(2*d{\usc}(mu,nu)*(g{\usc}(1,q1) - g{\usc}(1,q2)} \\
		& \phantom{=} \texttt{- 2*m*gi{\usc}(1)) - (q1(mu) - q2(mu))*g{\usc}(1,nu)} \\
		& \phantom{=} \texttt{- (q1(nu) - q2(nu))*g{\usc}(1,mu))}
	\end{split} \\
	\begin{split}\label{eqn:fr-v-gravitongluontriple-form}
		\Phi \left ( \scriptstyle \mu \nu \tcgreen{v-gravitongluontriple}^{\scriptstyle 2}_{\scriptstyle 1} \rule{0pt}{5ex} \right ) & = \texttt{i{\usc}*vkappa/2*d{\usc}(a1,a2)*(q1.q2*(d{\usc}(mu,nu)*d{\usc}(r1,r2)} \\
		& \phantom{=} \texttt{- d{\usc}(mu,r1)*d{\usc}(nu,r2) - d{\usc}(mu,r2)*d{\usc}(nu,r1))} \\
		& \phantom{=} \texttt{- d{\usc}(mu,nu)*q1(r2)*q2(r1) - d{\usc}(r1,r2)*(q1(mu)*q2(nu)} \\
		& \phantom{=} \texttt{+ q2(mu)*q1(nu)) + q1(r2)*(d{\usc}(mu,r1)*q2(nu)} \\
		& \phantom{=} \texttt{+ d{\usc}(nu,r1)*q2(mu)) + q2(r1)*(d{\usc}(mu,r2)*q1(nu)} \\
		& \phantom{=} \texttt{+ d{\usc}(nu,r2)*q1(mu)) + 1/xi*d{\usc}(mu,nu)*(q1(r1)*q2(r2)} \\
		& \phantom{=} \texttt{+ q1(r1)*q1(r2) + q2(r1)*q2(r2))} \\
		& \phantom{=} \texttt{- 1/xi*q1(r1)*(q1(nu)*d{\usc}(mu,r2) + q1(mu)*d{\usc}(nu,r2))} \\
		& \phantom{=} \texttt{- 1/xi*q2(r2)*(q2(nu)*d{\usc}(mu,r1) + q2(mu)*d{\usc}(nu,r1)))}
	\end{split} \\
	\begin{split} \label{eqn:fr-v-gravitongluonghosttriple-form}
		\Phi \left ( \scriptstyle \mu \nu \tcgreen{v-gravitongluonghosttriple}^{\scriptstyle q_2, b}_{\scriptstyle q_1, a} \right ) & = \texttt{i{\usc}*vkappa/(2*xi)*d{\usc}(a,b)*(q1.q2*d{\usc}(mu,nu)} \\
		& \phantom{=} \texttt{- q1(mu)*q2(nu) - q2(mu)*q1(nu))}
	\end{split} \\
	\begin{split} \label{eqn:fr-v-gravitontriple-form}
		\Phi \left ( \scriptstyle 1 \tcgreen{v-gravitontriple}^{\scriptstyle 3}_{\scriptstyle 2} \rule{0pt}{5ex} \right ) & = \texttt{-i{\usc}*vkappa/32} \sum_{\substack{\operatorname{Perm} \left ( 1, 2, 3 \right ) \\ \mu_i \leftrightarrow \nu_i}} \\
		& \phantom{=} \qquad \quad \texttt{(2*p1(n2)*p2(n3)*d\usc(m1,m2)*d\usc(m3,n1)} \\
		& \phantom{=} \qquad \quad \texttt{+ p1(n2)*p2(n1)*d\usc(m1,m3)*d\usc(m2,n3)} \\
		& \phantom{=} \qquad \quad \texttt{- p1(n1)*p2(n3)*d\usc(m1,m3)*d\usc(m2,n2)} \\
		& \phantom{=} \qquad \quad \texttt{- p1(n3)*p2(n1)*d\usc(m1,m3)*d\usc(m2,n2)} \\
		& \phantom{=} \qquad \quad \texttt{- p1(m2)*p2(n2)*d\usc(m1,m3)*d\usc(n1,n3)} \\
		& \phantom{=} \qquad \quad \texttt{+ 1/2*p1(m3)*p2(n3)*d\usc(m1,n1)*d\usc(m2,n2)} \\
		& \phantom{=} \qquad \quad \texttt{+ 1/2*p1(m2)*p2(n2)*d\usc(m1,n1)*d\usc(m3,n3)} \\
		& \phantom{=} \qquad \quad \texttt{- 1/2*p1(n2)*p2(n1)*d\usc(m1,m2)*d\usc(m3,n3)} \\
		& \phantom{=} \qquad \quad \texttt{- 1/2*p1(m3)*p2(n3)*d\usc(m1,m2)*d\usc(n1,n2)} \\
		& \phantom{=} \qquad \quad \texttt{+ p1.p2*d\usc(m1,m3)*d\usc(m2,n2)*d\usc(n1,n3)} \\
		& \phantom{=} \qquad \quad \texttt{- p1.p2*d\usc(m1,m2)*d\usc(m3,n1)*d\usc(n2,n3)} \\
		& \phantom{=} \qquad \quad \texttt{+ 1/4*p1.p2*d\usc(m1,m2)*d\usc(m3,n3)*d\usc(n1,n2)} \\
		& \phantom{=} \qquad \quad \texttt{- 1/4*p1.p2*d\usc(m1,n1)*d\usc(m2,n2)*d\usc(m3,n3))}
	\end{split} \\
	\begin{split} \label{eqn:fr-v-gravitonghosttriple-form}
		\Phi \left ( \scriptstyle \mu \nu \tcgreen{v-gravitonghosttriple}^{\scriptstyle 2}_{\scriptstyle 1} \right ) & = \texttt{i\usc*vkappa/8*(- d\usc(mu,nu)*q1(s1)*q1(s2)} \\
		& \phantom{=} \mkern-2mu \texttt{- 2*d\usc(mu,nu)*q1(s1)*q2(s2) - d\usc(mu,nu)*q2(s1)*q2(s2)} \\
		& \phantom{=} \mkern-2mu \texttt{- d\usc(mu,s1)*d\usc(nu,s2)*q1.q1 - 2*d\usc(mu,s1)*d\usc(nu,s2)*q1.q2} \\
		& \phantom{=} \mkern-2mu \texttt{- d\usc(mu,s1)*d\usc(nu,s2)*q2.q2 + d\usc(mu,s1)*q1(nu)*q1(s2)} \\
		& \phantom{=} \mkern-2mu \texttt{+ 2*d\usc(mu,s1)*q1(nu)*q2(s2) - d\usc(mu,s1)*q1(s2)*q2(nu)} \\
		& \phantom{=} \mkern-2mu \texttt{- d\usc(mu,s2)*d\usc(nu,s1)*q1.q1 - 2*d\usc(mu,s2)*d\usc(nu,s1)*q1.q2} \\
		& \phantom{=} \mkern-2mu \texttt{- d\usc(mu,s2)*d\usc(nu,s1)*q2.q2 - d\usc(mu,s2)*q1(nu)*q2(s1)} \\
		& \phantom{=} \mkern-2mu \texttt{+ 2*d\usc(mu,s2)*q1(s1)*q2(nu) + d\usc(mu,s2)*q2(nu)*q2(s1)} \\
		& \phantom{=} \mkern-2mu \texttt{+ d\usc(nu,s1)*q1(mu)*q1(s2) + 2*d\usc(nu,s1)*q1(mu)*q2(s2)} \\
		& \phantom{=} \mkern-2mu \texttt{- d\usc(nu,s1)*q1(s2)*q2(mu) - d\usc(nu,s2)*q1(mu)*q2(s1)} \\
		& \phantom{=} \mkern-2mu \texttt{+ 2*d\usc(nu,s2)*q1(s1)*q2(mu) + d\usc(nu,s2)*q2(mu)*q2(s1))}
	\end{split}
\end{align}
}%

\end{appendix}

\bibliography{References}
\bibliographystyle{babunsrt}

\end{document}